\documentclass{aa}
\usepackage{graphicx}
\usepackage[varg]{txfonts}
\usepackage{xcolor}
\usepackage{amsmath}
\usepackage{verbatim}
\usepackage{textcomp, gensymb}
\usepackage{acronym}
\usepackage{natbib}
\usepackage[colorlinks=true,citecolor=blue]{hyperref} 
\usepackage{tablefootnote}
\usepackage{multirow}
\usepackage{tabularx,booktabs}
\usepackage{subfigure}

\newcolumntype{C}{>{\centering\arraybackslash}X}
\newcolumntype{L}{>{\raggedright\arraybackslash}X}
\newcolumntype{R}{>{\raggedleft\arraybackslash}X}

\newcommand{\sidekicks}{\textsc{SideKicks.jl}}
\newcommand{\juliaCode}{\textsc{Julia}}
\newcommand{\Msun}{\ensuremath{\xspace M_{\odot}}\xspace}
\newcommand{\Rsun}{\ensuremath{\xspace R_{\odot}}\xspace}
\newcommand{\kms}{\ensuremath{\xspace\mathrm{km}\,\mathrm{s}^{-1}}\xspace}
\newcommand{\kpc}{\ensuremath{\xspace\mathrm{kpc}}\xspace}
\newcommand{\mas}{\ensuremath{\xspace\mathrm{mas}}\xspace}
\newcommand{\masyr}{\ensuremath{\xspace\mathrm{mas}\;\mathrm{yr}^{-1}}\xspace}
\newcommand{\curlyL}{\ensuremath{\mathcal{L}}\xspace}
\newcommand{\curlyN}{\ensuremath{\mathcal{N}}\xspace}
\newcommand{\curlyZ}{\ensuremath{\mathcal{Z}}\xspace}
\newcommand{\dLMC}{\ensuremath{d_\mathrm{LMC}}\xspace}
\newcommand{\mOnei}{\ensuremath{m_{\mathrm{1,i}}}\xspace}
\newcommand{\mOnef}{\ensuremath{m_{\mathrm{1,f}}}\xspace}
\newcommand{\mTwoi}{\ensuremath{m_{\mathrm{2,i}}}\xspace}
\newcommand{\mTwof}{\ensuremath{m_{\mathrm{2,f}}}\xspace}
\newcommand{\Mi}{\ensuremath{M_{\mathrm{i}}}\xspace}
\newcommand{\Mf}{\ensuremath{M_{\mathrm{f}}}\xspace}
\newcommand{\ai}{\ensuremath{a_{\mathrm{i}}}\xspace}
\newcommand{\af}{\ensuremath{a_{\mathrm{f}}}\xspace}

\newcommand{\Peri}{\ensuremath{P_{\mathrm{i}}}\xspace}
\newcommand{\Perf}{\ensuremath{P_{\mathrm{f}}}\xspace}
\newcommand{\PerCirc}{\ensuremath{P_{\mathrm{circ}}}\xspace}
\newcommand{\ei}{\ensuremath{e_{\mathrm{i}}}\xspace}
\newcommand{\ef}{\ensuremath{e_{\mathrm{f}}}\xspace}
\newcommand{\muenv}{\ensuremath{\vec{\mu}_{\mathrm{env}}}\xspace}
\newcommand{\sigmaenv}{\ensuremath{\sigma_{\mathrm{env}}}\xspace}
\newcommand{\venvk}{\ensuremath{\vec{v}_{\mathrm{env,k}}}\xspace}

\newcommand{\deltavr}{\ensuremath{\Delta v_\mathrm{r}}\xspace}
\newcommand{\deltavra}{\ensuremath{\Delta v_{\alpha}}\xspace}
\newcommand{\deltavdec}{\ensuremath{\Delta v_{\delta}}\xspace}
\newcommand{\deltav}{\ensuremath{\Delta v}\xspace}
\newcommand{\vri}{\ensuremath{v_{\mathrm{r,i}}}\xspace}
\newcommand{\vrai}{\ensuremath{v_{\alpha,\mathrm{i}}}\xspace}
\newcommand{\vdeci}{\ensuremath{v_{\delta,\mathrm{i}}}\xspace}
\newcommand{\viVec}{\ensuremath{\vec{v}_\mathrm{i}}\xspace}
\newcommand{\vfVec}{\ensuremath{\vec{v}_\mathrm{f}}\xspace}
\newcommand{\deltavVec}{\ensuremath{\Delta\vec{v}}\xspace}
\newcommand{\vrf}{\ensuremath{v_{\mathrm{r,f}}}\xspace}
\newcommand{\vraf}{\ensuremath{v_{\alpha,\mathrm{f}}}\xspace}
\newcommand{\vdecf}{\ensuremath{v_{\delta,\mathrm{f}}}\xspace}
\newcommand{\venvr}{\ensuremath{V_\mathrm{env,r}}\xspace}
\newcommand{\venvra}{\ensuremath{V_{\mathrm{env},\alpha}}\xspace}
\newcommand{\venvdec}{\ensuremath{V_{\mathrm{env},\delta}}\xspace}
\newcommand{\venvVec}{\ensuremath{\vec{V}_\mathrm{env}}\xspace}
\newcommand{\modelS}{\ensuremath{\mathrm{S}}\xspace}
\newcommand{\modelSVr}{\ensuremath{\mathrm{SV}_\mathrm{R}}\xspace}
\newcommand{\modelSVt}{\ensuremath{\mathrm{SV}_\mathrm{T}}\xspace}
\newcommand{\modelSVthree}{\ensuremath{\mathrm{SV}_\mathrm{3}}\xspace}
\newcommand{\modelSO}{\ensuremath{\mathrm{SO}}\xspace}
\newcommand{\modelSOVr}{\ensuremath{\mathrm{SOV}_\mathrm{R}}\xspace}
\newcommand{\modelSOVt}{\ensuremath{\mathrm{SOV}_\mathrm{T}}\xspace}
\newcommand{\modelSOVthree}{\ensuremath{\mathrm{SOV}_\mathrm{3}}\xspace}
\newcommand{\ehatx}{\ensuremath{\hat{e}_\mathrm{x}}\xspace}
\newcommand{\ehaty}{\ensuremath{\hat{e}_\mathrm{y}}\xspace}
\newcommand{\ehatz}{\ensuremath{\hat{e}_\mathrm{z}}\xspace}
\newcommand{\ehatPar}{\ensuremath{\hat{e}_\parallel}\xspace}
\newcommand{\ehatPrp}{\ensuremath{\hat{e}_\perp}\xspace}

\newcommand{\gnui}{\ensuremath{g_{\nu,\mathrm{i}}}\xspace}
\newcommand{\hnui}{\ensuremath{h_{\nu,\mathrm{i}}}\xspace}
\newcommand{\jnui}{\ensuremath{j_{\nu,\mathrm{i}}}\xspace}
\newcommand{\jnu}{\ensuremath{j_{\nu}}\xspace}
\newcommand{\omegai}{\ensuremath{\omega_{\mathrm{i}}}\xspace}
\newcommand{\omegaf}{\ensuremath{\omega_{\mathrm{f}}}\xspace}
\newcommand{\Omegai}{\ensuremath{\Omega_{\mathrm{i}}}\xspace}
\newcommand{\Omegaf}{\ensuremath{\Omega_{\mathrm{f}}}\xspace}
\newcommand{\inci}{\ensuremath{i_{\mathrm{i}}}\xspace}
\newcommand{\incf}{\ensuremath{i_{\mathrm{f}}}\xspace}
\newcommand{\inc}{\ensuremath{i}\xspace}
\newcommand{\KOne}{\ensuremath{K_1}\xspace}

\newcommand{\nui}{\ensuremath{\nu_{\mathrm{i}}}\xspace}
\newcommand{\nuf}{\ensuremath{\nu_{\mathrm{f}}}\xspace}
\newcommand{\fnui}{\ensuremath{f_{\nu,\mathrm{i}}}\xspace}
\newcommand{\fnuf}{\ensuremath{f_{\nu,\mathrm{f}}}\xspace}
\newcommand{\vkick}{\ensuremath{v_\mathrm{k}}\xspace}
\newcommand{\vkickVec}{\ensuremath{\vec{v}_\mathrm{k}}\xspace}
\newcommand{\vkickHat}{\ensuremath{\hat{v}_\mathrm{k}}\xspace}
\newcommand{\dMOne}{\ensuremath{\Delta m_1}\xspace}
\newcommand{\dMTwo}{\ensuremath{\Delta m_2}\xspace}
\newcommand{\fdMTwo}{\ensuremath{f_{\Delta M}}\xspace}
\newcommand{\vimp}{\ensuremath{v_\mathrm{imp}}\xspace}
\newcommand{\vOnei}{\ensuremath{\vec{v}_\mathrm{1,i}}\xspace}
\newcommand{\vOnef}{\ensuremath{\vec{v}_\mathrm{1,f}}\xspace}
\newcommand{\vTwoi}{\ensuremath{\vec{v}_\mathrm{2,i}}\xspace}
\newcommand{\vTwof}{\ensuremath{\vec{v}_\mathrm{2,f}}\xspace}
\newcommand{\vreli}{\ensuremath{v_\mathrm{rel,i}}\xspace}

\newcommand{\vCMf}{\ensuremath{\vec{v}_\mathrm{CM,f}}\xspace}
\newcommand{\vCMOnef}{\ensuremath{\vec{v}_\mathrm{CM,1,f}}\xspace}
\newcommand{\vCMTwof}{\ensuremath{\vec{v}_\mathrm{CM,2,f}}\xspace}
\newcommand{\rCMOnef}{\ensuremath{\vec{r}_\mathrm{CM,1,f}}\xspace}
\newcommand{\rCMTwof}{\ensuremath{\vec{r}_\mathrm{CM,2,f}}\xspace}
\newcommand{\vCMPar}{\ensuremath{v_{\mathrm{CM},\parallel}}\xspace}
\newcommand{\vCMPrp}{\ensuremath{v_{\mathrm{CM},\perp}}\xspace}
\newcommand{\vCMz}{\ensuremath{v_{\mathrm{CM,z}}}\xspace}
\newcommand{\Ef}{\ensuremath{E_{\mathrm{f}}}\xspace}
\newcommand{\Lf}{\ensuremath{L_{\mathrm{f}}}\xspace}
\newcommand{\LfVec}{\ensuremath{\vec{L}_{\mathrm{f}}}\xspace}
\newcommand{\LfHat}{\ensuremath{\hat{L}_{\mathrm{f}}}\xspace}
\newcommand{\OmegafHat}{\ensuremath{\hat{\Omega}_{\mathrm{f}}}\xspace}
\newcommand{\bPar}{\ensuremath{b_\parallel}\xspace}
\newcommand{\bPrp}{\ensuremath{b_\perp}\xspace}

\newcommand{\bW}{\ensuremath{b_\mathrm{W}}\xspace}
\newcommand{\bN}{\ensuremath{b_\mathrm{N}}\xspace}
\newcommand{\bO}{\ensuremath{b_\mathrm{O}}\xspace}

\newcommand{\pax}{\ensuremath{\pi}\xspace}
\newcommand{\pmra}{\ensuremath{\mu_{\alpha}}\xspace}

\newcommand{\pmdec}{\ensuremath{\mu_{\delta}}\xspace}
\newcommand{\gaia}{\emph{Gaia}\xspace}
\newcommand{\bloem}{\ac{BLOeM}\xspace}

\acrodef{BH}{black hole}
\acrodef{BHB}{black hole binary}
\acrodefplural{BHB}{black hole binaries}
\acrodef{SN}{supernova}
\acrodefplural{SN}[SNe]{supernovae}
\acrodef{CC}{core-collapse}
\acrodef{NS}{neutron star}
\acrodef{XRB}{X-ray binary}
\acrodefplural{XRB}{X-ray binaries}
\acrodef{RV}{radial velocity}
\acrodef{WR}{Wolf-Rayet}
\acrodef{BLOeM}{Binarity at LOw Metallicity}
\acrodef{DR4}{data release 4}
\acrodef{CI}{confidence interval}
\newcommand{\ibhb}{inert \ac{BHB}\xspace}

\newcommand{\ibhbs}{inert \acp{BHB}\xspace}
\newcommand{\Ibhbs}{Inert \acp{BHB}\xspace}

\usepackage{tikz,xcolor,hyperref}
\definecolor{lime}{HTML}{A6CE39}
\DeclareRobustCommand{\orcidicon}{\hspace{-1mm}
	\begin{tikzpicture}
	\draw[lime, fill=lime] (0,0)
	circle [radius=0.16]
	node[white] {{\fontfamily{qag}\selectfont \tiny \,ID}};
	\draw[white, fill=white] (-0.0525,0.095)
	circle [radius=0.007];
	\end{tikzpicture}
	\hspace{-3mm}
}
\newcommand{\orcid}[1]{\href{https://orcid.org/#1}{\orcidicon}}

\begin{document} 

\title{Binarity at LOw Metallicity (BLOeM):}
\subtitle{Bayesian inference of natal kicks from inert black hole binaries}
    \author{ R.\ Willcox 
      \orcid{0000-0003-1817-3586} 
      \inst{\ref{inst:kul},\ref{inst:lgi}}
      \thanks{\href{mailto:reinhold.willcox@kuleuven.be}{reinhold.willcox@kuleuven.be}} 
    \and  P.\ Marchant\orcid{0000-0002-0338-8181} 
      \inst{\ref{inst:gent},\ref{inst:kul}}
    \and A. Vigna-G\'omez\orcid{0000-0003-1817-3586}
      \inst{\ref{inst:mpa}}
    \and H. Sana 
      \inst{\ref{inst:kul},\ref{inst:lgi}}
    \and J. Bodensteiner\orcid{0000-0002-9552-7010}
      \inst{\ref{inst:UoAms},\ref{inst:eso}}
    \and K. Deshmukh\orcid{0000-0001-5253-3480}
      \inst{\ref{inst:kul},\ref{inst:lgi}}
    \and M. Esseldeurs\orcid{0000-0002-4650-6029} 
      \inst{\ref{inst:kul}}
    \and M. Fabry
      \inst{\ref{inst:villa},\ref{inst:kul}}
    \and V. H\'enault-Brunet\orcid{0000-0003-2927-5465}
      \inst{\ref{inst:smu}}
    \and S. Janssens\orcid{0000-0002-9758-4289}
      \inst{\ref{inst:utokyo}}
    \and L. Mahy\orcid{0000-0003-0688-7987}
      \inst{\ref{inst:rob}}
    \and L. Patrick 
      \inst{\ref{inst:cab}}
    \and D. Pauli\orcid{0000-0002-5453-2788}
      \inst{\ref{inst:kul}}
    \and M. Renzo\orcid{0000-0002-6718-9472}
      \inst{\ref{inst:ariz}}
    \and A. A. C. Sander\orcid{0000-0002-2090-9751}
      \inst{\ref{inst:ari}}
    \and T. Shenar
      \inst{\ref{inst:telaviv}}
    \and L.~A.~C.~van~Son\orcid{0000-0001-5484-4987}
      \inst{\ref{inst:cca},\ref{inst:princeton}}
    \and M. Stoop\orcid{0000-0003-4723-0447} 
      \inst{\ref{inst:UoAms}}
    }

    \institute{
    {Institute of Astronomy, KU Leuven, Celestijnenlaan 200D, 3001 Leuven, Belgium\label{inst:kul}} 
    \and {Leuven Gravity Institute, KU Leuven, Celestijnenlaan 200D, box 2415, 3001 Leuven, Belgium \label{inst:lgi}}
    \and {Sterrenkundig Observatorium, Universiteit Gent, Krijgslaan 281 S9, B-9000 Gent, Belgium\label{inst:gent}}
    \and {Max-Planck-Institute for Astrophysics, Karl-Schwarzschild-Strasse 1, 85748 Garching, Germany\label{inst:mpa}}
    \and {Anton Pannekoek Institute for Astronomy, University of Amsterdam, Science Park 904, 1098 XH Amsterdam, The Netherlands \label{inst:UoAms}}
    \and {ESO - European Southern Observatory, Karl-Schwarzschild-Strasse 2, 85748 Garching bei München, Germany\label{inst:eso}}
    \and {Department of Astrophysics and Planetary Science, Villanova University, 800 E Lancaster Ave., Villanvona, PA 19085\label{inst:villa}}
    \and {Department of Astronomy and Physics, Saint Mary's University, 923 Robie Street, Halifax, B3H 3C3, Canada\label{inst:smu}}
    \and {Research Center for the Early Universe, Graduate School of Science, University of Tokyo, Bunkyo, Tokyo 113-0033, Japan. \label{inst:utokyo}}
    \and {Royal Observatory of Belgium, Department of Astronomy and Astrophysics, Avenue Circulaire/Ringlaan 3, B-1180, Brussels, Belgium \label{inst:rob}}
    \and {Centro de Astrobiologia (CAB), CSIC-INTA, Carretera de Ajalvir km4, 28850 Torrejón de Ardoz, Madrid, Spain \label{inst:cab}}
    \and {University of Arizona, Department of Astronomy \& Steward Observatory, 933 N. Cherry Ave., Tucson, AZ 85721, USA\label{inst:ariz}}
    \and {Zentrum f\"ur Astronomie der Universit\"at Heidelberg, Astronomisches Rechen-Institut, M\"onchhofstr. 12-14, 69120 Heidelberg, Germany\label{inst:ari}} 
    \and {The School of Physics and Astronomy, Tel Aviv University, Tel Aviv 6997801, Israel\label{inst:telaviv}}
    \and {Center for Computational Astrophysics, Flatiron Institute, New York, NY 10010, USA\label{inst:cca} }
    \and {Department of Astrophysical Sciences, Princeton University, 4 Ivy Lane, Princeton, NJ 08544, USA\label{inst:princeton}}
    }
    
   \date{Received XXX; accepted XXX}

 
  \abstract
   {
    The emerging population of inert black hole binaries (BHBs) provides a unique opportunity to constrain black hole (BH) formation physics. These systems are composed of a stellar-mass BH in a wide orbit around a non-degenerate star with no observed X-ray emission. Inert BHBs allow for narrow constraints to be inferred on the natal kick and mass loss during BH-forming core-collapse events. 
   }
   {
   In anticipation of the upcoming BLOeM survey, we aim to provide tight constraints on BH natal kicks by exploiting the full parameter space obtained from combined spectroscopic and astrometric data to characterize the orbits of inert BHBs. Multi-epoch spectroscopy from the BLOeM project will provide measurements of periods, eccentricities, and radial velocities for inert BHBs in the SMC, which complements \gaia astrometric observations of proper motions. 
   } 
   {
    We present a Bayesian parameter estimation framework to infer natal kicks and mass loss during core-collapse from inert BHBs, accounting for all available observables, including the systemic velocity and its orientation relative to the orbital plane. 
    The framework further allows for circumstances when some of the observables are unavailable, such as for the distant BLOeM sources which preclude resolved orbits. 
    This method is implemented in a publicly available open source package, \sidekicks.
   }
   {
   With our new framework, we are able to distinguish between BH formation channels, even in the absence of a resolved orbit. 
   In cases when the pre-explosion orbit can be assumed to be circular, we precisely recover the parameters of the core-collapse, highlighting the importance of understanding the eccentricity landscape of pre-explosion binaries, both theoretically and observationally. 
   Treating the near-circular, inert BHB, VFTS 243, as a representative of the anticipated BLOeM systems, we constrain the natal kick to $\lesssim27~\kms$ and the mass loss to $\lesssim2.9~\Msun$ within a 90\% credible interval. 
   }
   {}


   \titlerunning{Natal Kick Inference from Inert BH Binaries}
   \authorrunning{R. Willcox et al.}
   \maketitle
%

\section{Introduction}

Massive stars (those with initial mass $\gtrsim 8\;\Msun$) end their lives in a \ac{CC} event, possibly with an associated bright \ac{SN} explosion or gamma-ray burst, resulting in a compact object: either a \ac{NS} or a \ac{BH}.
The majority of massive stars are now known to be born in binaries or higher multiplicity systems, with the multiplicity fraction of stars with $M\gtrsim10~\Msun$ approaching 100\% \citep{Sana_etal.2012_BinaryInteractionDominates, Moe_DiStefano.2017_MindYourPs, Offner_etal.2023_OriginEvolutionMultiple}.  During compact object formation, asymmetries in the \ac{SN} ejecta can drive a momentum recoil, or ``natal kick'', in the remnant, potentially resulting in high velocity compact objects \citep{Katz.1975_TwoKindsStellar, Brisken_etal.2003_ProperMotionMeasurementsVla, Hobbs_etal.2005_StatisticalStudy233, Verbunt_etal.2018_ObservedVelocityDistribution}. In binary systems, even symmetric mass loss alone with no additional natal kick (often referred to as a ``Blaauw kick'') can significantly modify the orbit or result in isolated compact objects \citep{Blaauw.1961_OriginBtypeStars}. 

Presently, the kicks that stellar-mass \acp{BH} attain upon birth represent a major source of uncertainty in stellar evolution modeling, with implications for the formation of stellar binaries containing \acp{BH} \citep{Mandel.2016_EstimatesBlackholeNatal}, retention of \acp{BH} in globular clusters \citep{Antonini_Gieles.2020_PopulationSynthesisBlack}, the velocity of dark lenses \citep{Wyrzykowski_Mandel.2020_ConstrainingMassesMicrolensing}, the mass distribution of massive runaway stars \citep{Dray_etal.2006_SupernovaRunawayStars, Renzo_etal.2019_MassiveRunawayWalkaway}, and the properties of merging binary \acp{BH} \citep{Callister_etal.2021_StateFieldBinary, Broekgaarden_etal.2022_ImpactMassiveBinary}.

Estimating \ac{BH} natal kicks is challenging due to the inherent difficulty of detecting \acp{BH} in general. Most stellar mass \acp{BH} are discovered in binary systems, often as low- or high-mass \acp{XRB}.  Low-mass \acp{XRB} may be very old, such that estimating the natal kick requires tracing the system back through its Galactic orbit \citep{Nagarajan_El-Badry.2025_MixedOriginsStrong}, while assuming that the binary formed near the disk and was unperturbed by third-body encounters in the interim.  Furthermore, that method can only constrain the full systemic velocity following the \ac{CC}, and cannot disentangle the effects of the natal kick, mass loss, and any pre-collapse systemic velocity \citep{Atri_etal.2019_PotentialKickVelocity}. High-mass \acp{XRB}, by contrast, are constrained by the age of the visible companion to be relatively young, and are thus more likely to have a velocity that reflects their speed immediately following the \ac{CC} event, although rejuvenation during mass accretion may modify this age constraint.  \citet{Mirabel_Rodrigues.2003_FormationBlackHole} used the proper motion of the high-mass \ac{XRB} Cyg X-1 to argue that its low systemic velocity, relative to its presumed birth association, Cygnus OB3, was indicative of a small natal kick and minimal mass loss.  Since larger kicks are more likely to disrupt binaries, \acp{BH} observed in binaries are evidence that at least some \acp{BH} form with weak kicks. Consequently, kick distributions estimated from \acp{XRB} are intrinsically biased against larger kicks  \citep{Repetto_etal.2012_InvestigatingStellarmassBlack, Mandel.2016_EstimatesBlackholeNatal, Renzo_etal.2019_MassiveRunawayWalkaway, Andrews_Kalogera.2022_ConstrainingBlackHole, ODoherty_etal.2023_ObservationallyDerivedKick, Kimball_etal.2023_BlackHoleKicked, Burdge_etal.2024_BlackHoleLow, DashwoodBrown_etal.2024_NatalKickBlack}.

In addition to constraints from \acp{BH} born in isolated binaries, the existence of \acp{BH} in globular clusters (e.g. \citet{Baumgardt_etal.2019_NoEvidenceIntermediatemass, Weatherford_etal.2020_DynamicalSurveyStellarmass, Dickson_etal.2024_MultimassModellingMilky}) provides evidence that at least some \acp{BH} receive kicks that are less than the cluster escape velocity.  The central escape velocity of Milky Way globular clusters is typically on the order of $\sim20\kms$ at the present day, and it could have been higher by a factor of a few at the time of \ac{BH} formation due to clusters mass loss and expansion over time \citep{Baumgardt_Hilker.2018_CatalogueMassesStructural}. However, due to dynamical interactions within the cluster, the present-day velocity of a cluster \ac{BH} may be very different from its natal kick, so we cannot infer the natal kicks on a per system basis.

The mass distribution of runaway massive stars is also sensitive to \ac{BH} natal kicks. If \ac{BH} progenitors typically have more massive companions, then larger \ac{BH} kicks imply more massive runaways, and this can provide a statistical constraint on the kicks as a function of the initial mass ratio distribution, even in the absence of an observed core-collapse or \ac{BH} \citep{Renzo_etal.2019_MassiveRunawayWalkaway}. 

Studies of weak \ac{NS} natal kicks face similar challenges due to the dynamical effects in clusters and the evolutionary effects on the orbit of \acp{NS} in \ac{NS}-\acp{XRB} \citep{ODoherty_etal.2023_ObservationallyDerivedKick}. 
However, transverse velocity observations of hundreds of isolated radio pulsars provide evidence for typical \ac{NS} natal kicks following a Maxwellian distribution with argument $\sigma=265\;\kms$ \citep{Hobbs_etal.2005_StatisticalStudy233}. As the complement to \acp{NS} in intact binaries, isolated pulsars velocities are naturally biased towards stronger kicks, and different studies disagree on the magnitude and shape of the low velocity end of the \ac{NS} natal kick distribution \citep{Arzoumanian_etal.2002_VelocityDistributionIsolated, Podsiadlowski_etal.2005_NeutronStarBirthKicks, Igoshev_etal.2021_CombinedAnalysisNeutron, Willcox_etal.2021_ConstraintsWeakSupernova, Fortin_etal.2022_ConstraintsNeutronstarKicks, Zhao_etal.2023_EvidenceMassdependentPeculiar}.

Unlike radio pulsars, isolated \acp{BH} cannot be observed directly, although microlensing provides a method to infer their velocities. The only confirmed detection of a microlensed stellar-mass \ac{BH} with a velocity constraint is OGLE-2011-BLG-0462/MOA-2011-BLG-191  \citep{Sahu_etal.2022_IsolatedStellarmassBlack, Lam_etal.2022_IsolatedMassgapBlack}. Although the original nature and properties of OGLE-2011-BLG-0462/MOA-2011-BLG-191 were disputed \citep{Mroz_etal.2022_SystematicErrorsSource}, subsequent analysis has confidently placed the \ac{BH} mass between $\sim6-8~\Msun$ and the transverse velocity at $\lesssim 45~\kms$ \citep{Sahu_etal.2022_IsolatedStellarmassBlack,Mroz_etal.2022_SystematicErrorsSource}, with an estimated \ac{BH} natal kick of $\lesssim100~\kms$ \citep{Andrews_Kalogera.2022_ConstrainingBlackHole}. Based on the observed rate of microlensing events, \citet{Koshimoto_etal.2024_InfluenceBlackHole} estimate that the average kick velocity of isolated \acp{BH} should be $\lesssim 100~\kms$, however this analysis did not account for the likelihood that the \acp{BH} observed in microlensing events are most likely of binary origin \citep{Vigna-Gomez_Ramirez-Ruiz.2023_BinaryOriginFirst}.

A promising yet computationally costly alternative to study \ac{BH} natal kicks comes from 3D \ac{SN} simulations \citep{Janka_etal.2016_PhysicsCoreCollapseSupernovae, Muller.2020_HydrodynamicsCorecollapseSupernovae, Mezzacappa.2023_RealisticModelsCore, Wang_etal.2022_EssentialCharacterNeutrino, Janka_etal.2022_SupernovaFallbackOrigin}.  While some recent studies have argued that most \acp{BH} receive kicks $<100~\kms$ \citep{Coleman_Burrows.2022_KicksInducedSpins}, others have shown that some newly formed \acp{BH} may attain kicks larger than $1000~\kms$, at least at some point during the explosion \citep{Burrows_etal.2024_PhysicalCorrelationsPredictions, Janka_Kresse.2024_InterplayNeutrinoKicks}. Such extreme values are expected to reduce with the accretion of fallback material from the envelope \citep{Schroder_etal.2018_BlackHoleFormation, Chan_etal.2020_ImpactFallbackCompact}, however simulating for long enough to resolve these later phases is just at the edge of the computational capabilities of current simulations, and systematic trends are presently out of reach.  As such, there is no clear consensus from 3D modeling on the magnitude of \ac{BH} kicks and their dependence on the properties of the progenitor. 

Recent discoveries of a new population of \ibhbs provide a promising new window into \ac{BH} formation and direct constraints on natal kicks and mass loss \citep{Shenar_etal.2022_XrayQuietBlack}. Also referred to as ``detached'' or ``dormant'', \ibhbs are composed of a \ac{BH} and a non-degenerate star, with no observed X-ray emission or signatures of an accretion disk; the term ``inert'' distinguishes these from quiescent \acp{XRB}, which fluctuate between X-ray emitting and X-ray quiet (see \citealt{Hirai_Mandel.2021_ObservableBlackHole, Sen_etal.2024_WhisperingDarkFaint} for discussions).  The persistent lack of X-rays characterizes these as wide, non-interacting binaries.  Some are detected spectroscopically, such as the binary in NGC 3201 \citep{Giesers_etal.2018_DetachedStellarmassBlack}, the Galactic binary HD 130298 \citep{Mahy_etal.2022_IdentifyingQuiescentCompact}, and the LMC binary VFTS 243 \citep{Shenar_etal.2022_XrayQuietBlack}. Others, such as the three \gaia \acp{BH}, are detected via astrometry and later confirmed spectroscopically as \ibhbs \citep{El-Badry_etal.2023_SunlikeStarOrbiting, Chakrabarti_etal.2023_NoninteractingGalacticBlack, El-Badry_etal.2023_RedGiantOrbiting, Tanikawa_etal.2023_SearchBlackHole, GaiaCollaboration_etal.2024_DiscoveryDormant33}. Additionally, several candidate \ibhbs in the Milky Way \citep{Mahy_etal.2022_IdentifyingQuiescentCompact, Banyard_etal.2023_SearchingCompactObjects},  NGC 3201 \citep{Giesers_etal.2019_StellarCensusGlobular}, and the LMC \citep{Shenar_etal.2022_XrayQuietBlack, Shenar_etal.2022_TarantulaMassiveBinary} warrant follow-up confirmation.

Notably, the companion masses to all three \gaia \acp{BH}, as well as to the NGC 3201 \ac{BH}, are $\lesssim1~\Msun$, suggesting that these systems, like the low-mass \acp{XRB}, may be very old, and their velocities today may deviate from their birth velocities due to motion through the Galactic potential and the increased likelihood for third-body interactions \citep{Rastello_etal.2023_DynamicalFormationGaia}. 
By contrast, the companion masses in both HD 130298 and VFTS 243 are $\sim25~\Msun$, providing an upper age limit of $\sim10$~Myr (assuming no rejuvenation), which severely limits the potential for perturbations to the orbit or systemic velocity.

Moreover, for all of the confirmed systems listed above, it is believed that the effect of tides on these orbits is currently negligible (although this is not a requirement to be considered inert).  Given the short ages and weak tidal interactions for HD 130298 and VFTS 243, we expect that the observed orbital configurations and velocities for both binaries have not changed meaningfully since the formation of the \ac{BH}. Furthermore, the relatively short periods, $\sim15~d$ and $\sim10$~d for HD 130298 and VFTS 243, respectively, strongly suggest that these binaries interacted via mass transfer prior to \ac{CC}.  Binary evolution models ubiquitously assume that mass transfer drives a binary to circularization, however that assumption has recently  received more attention both observationally and theoretically (see discussion in Sec.~\ref{sec:importance_preCC_eccentricity}).

VFTS 243, in addition, has a very small measured eccentricity, $e = 0.017 \pm 0.012$ (1$\sigma$ level). This supports the direct collapse \ac{BH} formation hypothesis, with negligible natal kick and mass loss \citep{Shenar_etal.2022_XrayQuietBlack, Stevance_etal.2022_VFTS243Predicted, Banagiri_etal.2023_DirectStatisticalConstraints, Vigna-Gomez_etal.2024_ConstraintsNeutrinoNatal}. However, no prior study has used the transverse velocity information or possible correlations between the system velocity vector and the orbital elements (specifically, the Keplerian angles).

Many more \ibhbs are anticipated to be detected in the near future. By the end of 2025, the \bloem campaign will complete its multi-epoch spectroscopy of massive stars in the SMC, including massive binaries containing \acp{BH}, providing unprecedented constraints on \ac{BH} formation at low metallicity \citep{Shenar_etal.2024_BinarityLOwMetallicity}. Additionally, the upcoming \gaia \ac{DR4} is expected to provide binary solutions to dozens or potentially hundreds of massive, Galactic \ibhb candidates around mid-2026 \citep{Karpov_Lipunov.2001_WhyWeSee, Mashian_Loeb.2017_HuntingBlackHoles, Breivik_etal.2017_RevealingBlackHoles, Janssens_etal.2023_DetectionSingledegenerateMassive}.  However, not all of the observed \ibhbs will be able to provide useful constraints on \ac{BH} formation and natal kicks.  Given the expected influx of \ibhbs over the next few years, and the variety of observing constraints that will be available, a more complete statistical treatment is warranted.

In this study, we present a Bayesian parameter inference framework to analyze natal kicks in binary star systems, specifically in the context of \ibhbs.  The method is unique as it incorporates all possible observational information, including the full 3D velocity of the system with respect to its orbital elements, and accounts for any pre-collapse orbital eccentricity. 

We present the methodology of our inference pipeline in Sec.~\ref{sec:methodology}, including the estimation of the systemic velocity prior to collapse. We provide injection tests to highlight the strength of the inference in a mock system with optimal measurement uncertainty. In Sec.~\ref{sec:injection_tests}, we focus on the ideal case when all orbital elements are known, as well as the more realistic cases when some observational constraints have not yet been measured or are physically unobtainable. In Sec.~\ref{sec:vfts243}, we apply this formalism to VFTS 243 as a benchmark system to facilitate comparison to previous studies and as a representative of the anticipated BLOeM systems. We discuss the impact and limitations of these results in Sec.~\ref{sec:discussion}, as well as the circumstances that merit targeted follow-up observations. We conclude in Sec.~\ref{sec:conclusions}.

\section{Methodology}\label{sec:methodology}

We introduce our Bayesian parameter inference framework, in which we model how the \ac{CC} of a massive star in a binary impacts the intrinsic and extrinsic elements of the binary orbit.
We focus particularly on the inference gains that come from adding new information when different types of observation are available, including spectroscopy, proper motions, resolution of the orbit, and velocity measurements of the \ibhb and its birth association.
We discuss under what conditions different observables are available and how these restrict the solution space for the pre-explosion properties. 
If the current orbit is completely characterized, including the direction of motion of the center of mass relative to the orientation of the binary, and if the orbit was circular prior to the collapse, we demonstrate that we can completely recover all of the \ac{CC} parameters and pre-collapse orbital properties.

To facilitate the inference, we present our custom open-source \juliaCode{} package \sidekicks{} \footnote{Sidekicks: Statistical Inference to DEtermine KICKS. Publicly available at \url{https://github.com/orlox/SideKicks.jl}} which uses a Markov Chain Monte Carlo (MCMC) sampler to explore the likelihoods from any available observed quantities and arbitrary priors on the parameters that describe the pre-collapse orbit and the \ac{CC} event.

\subsection{Parameters of the orbit and the collapse}
\label{sec:parameters_orbit_collapse}
 
The problem of forward modeling a \ac{CC} event in a binary composed of two point masses has been treated extensively in the past  \citep{Blaauw.1961_OriginBtypeStars, Boersma.1961_MathematicalTheoryTwobody, Hills.1983_EffectsSuddenMass, Brandt_Podsiadlowski.1995_EffectsHighvelocitySupernova, Kalogera.1996_OrbitalCharacteristicsBinary, Tauris_Takens.1998_RunawayVelocitiesStellar, Pfahl_etal.2002_ComprehensiveStudyNeutron}; the procedure is repeated in App.~\ref{appx:explosion_math}, extended to include the information about all six post-explosion orbital elements as well as the full 3D velocity of the system.
We build on the previous efforts by explicitly calculating the Keplerian angles and velocity vector of the post-explosion system in the observer's frame of reference.

In the most general case, the pre-collapse binary is defined by four intrinsic parameters -- the initial period $\Peri$, eccentricity $\ei$, and both component masses $\mOnei, \mTwoi$ -- and six extrinsic parameters -- the orientation of the binary and the velocity vector of the center of mass, both relative to the observer. 
Throughout this text, the subscript ``$\mathrm{i}$'' refers to initial, or pre-collapse, parameters, while the subscript ``$\mathrm{f}$'' distinguishes final, or post-\ac{CC}, parameters. 
Additionally, subscript ``$1$'' corresponds to the currently visible star, consistent with the notation provided from observations, while subscript ``$2$'' corresponds to the component which experiences the \ac{CC}.
The orientation is described by the longitude of the ascending node $\Omegai$ and the argument of periastron $\omegai$ (both defined in terms of star 1), and the the orbital inclination $\inci$.
We take the convention here that the ascending node corresponds to the intersection point of the orbit with the celestial plane at which the star is moving away from the observer.

Additionally, during the \ac{CC}, there may be mass loss $\dMTwo$ and a natal kick $\vkickVec$ from the \ac{CC} progenitor, with polar and azimuthal kick angles $\theta$ and $\phi$, respectively, about the pre-collapse orbital velocity vector (see Figs.~\ref{appx:orbitaxes} and \ref{fig:kick_angles} for diagrams depicting this configuration). In the most general case, there may also be mass loss $\dMOne$ from ablation on the companion, and an associated impact kick $\vimp$ directed away from the collapsing star, however we ignore these throughout (though see Sec.~\ref{sec:impact_on_primary} for a discussion). If the pre-collapse orbit is eccentric, we also define the true anomaly at the moment of explosion $\nui$.

\subsection{Observable quantities}
\label{sec:observable_quantities}

The parameters that can be constrained for a given binary depend on the types of observations that are available. We assume that the \ibhbs under consideration are all observed spectroscopically, as this is required to confirm the nature of the \ac{BH}. 
Spectroscopic binaries subdivide into SB2s, when both stellar components can be identified from their spectral line variations, and SB1s, when only one component is sufficiently luminous to provide identifiable spectral lines. 
\Ibhbs, which can only be SB1s given that the companion is a dark object, provide measurements for the final orbital period $\Perf$, eccentricity $\ef$, argument of periastron $\omegaf$, \ac{RV} semiamplitude $\KOne$ of the visible companion, and the system \ac{RV} \vrf. 

From \Perf and \KOne, we can derive the \ac{BH} mass function,
\begin{equation}\label{eq:mass_function}
f_1 = \frac{\mTwof^3 \sin^3\incf}{(\mOnef + \mTwof)^2} = \frac{\Perf \KOne^3}{2\pi G}\left(1-\ef^2\right)^{3/2},
\end{equation}
with \incf the orbital inclination and $G$ the gravitational constant, which provides a lower limit for the \ac{BH} mass \mTwof. 
However, \mOnef itself can only be estimated using stellar atmosphere modelling, which may involve additional uncertainties (these are discussed further in Sec.~\ref{sec:model_dependency_primary_mass}).

Astrometric observations are also crucial for full characterization of binary orbits. Sources with astrometrically resolved orbits of the visible primary provide additional constraints on the longitude of the ascending node \Omegaf (defined for the visible companion) and the inclination \incf. Unfortunately, for distant sources like the \bloem targets located in the SMC, astrometric solutions taken from current \gaia data cannot resolve the orbit. The solutions can, however, provide a correlated measurement of the source parallax \pax and proper motions \pmra and \pmdec, in the directions of right ascension (RA, $\alpha$) and declination (Dec, $\delta$), respectively. Since the parallax measurements are consistent with zero for extragalactic sources, some care has to be taken to extract an informative subset of these solutions; we discuss this in detail in Sec.~\ref{sec:birth_velocity}. From the parallax and proper motion, we can derive the transverse systemic velocity components \vraf and \vdecf.


In all cases, the radial and transverse systemic velocities are derived relative to the observer, and so these are uninformative if the pre-collapse systemic velocity is not known.  In lieu of a method to measure the pre-collapse velocity directly, we assume that the binary was born with a velocity consistent with the dispersion of its birth association, although the identification of such an association may be non-trivial. We further assume that the velocity of the \ibhb of interest has not changed significantly in the time since the birth of the \ac{BH}, which we account for by limiting our analysis to \ibhbs where the visible component is a massive star. The short lifetime of the massive star provides an upper limit on the distance through which the binary could have traveled. 

A given velocity component for an \ibhb is only included in the analysis if the pre-collapse velocity in the same direction can be estimated from the birth association.
For example, if an observed binary has all three velocity components measured, but the reference binaries in the birth association have only astrometric proper motion measurements, then only the transverse velocity components are used and the systemic \ac{RV} information is omitted.  This is discussed in detail in Sec.~\ref{sec:birth_velocity}.

\subsection{Bayesian inference and MCMC}
\label{sec:bayesian_inference_and_mcmc}

In our Bayesian formalism, the parameters $\vec{\theta}$ that define the \ac{CC} and the pre-collapse orbit are described by the prior distributions 
$\vec{\pi}(\vec{\theta})$, while the measurements of the observations $\vec{d}$ for a given system are given by the likelihood $\vec{\curlyL}(\vec{d}|\vec{\theta})$, where $\vec{\theta}$ is distinct to the natal kick angle $\theta$. 
The posteriors $\vec{P}(\vec{\theta}|\vec{d})$ are thus given by

\begin{equation}
\centering
    \vec{P}(\vec{\theta}|\vec{d}) = \frac{ \vec{\curlyL}(\vec{d}|\vec{\theta}) \vec{\pi}(\vec{\theta}) }{\curlyZ(\vec{d})},
\end{equation}
where $\curlyZ(\vec{d})$ is the Bayesian evidence.

The posterior is estimated using an MCMC sampler. Within \sidekicks{}, we opt to use the \textsc{No-U-Turn Sampler (NUTS)} algorithm \citep{Homan_Gelman.2014_NoUturnSamplerAdaptively}, as implemented in the \juliaCode{} package \textsc{Turing.jl} \citep{Ge_etal.2018_TuringLanguageFlexible}.

\subsubsection{Priors}
\label{sec:priors}



For the results presented in this work, the priors $\vec{\pi}(\vec{\theta})$ are chosen to be broad and uninformative, although \sidekicks{} allows for any arbitrary definition of priors.  For the initial mass of the visible components \mOnei, we draw from a broad log-uniform distribution, \mbox{$\pi(\log_{10}(\mOnei/\Msun)) = U(0.1, 3)$}, and identically for the \ac{BH} progenitor \mTwoi. The initial period is drawn log uniformly, \mbox{$\pi(\log_{10}(\Peri/\mathrm{d})) = U(-1, 3)$}.  The initial eccentricity prior is either taken to be completely agnostic, \mbox{$\pi(\ei) = U(0,1)$}, or is fixed at 0 in cases where the initial orbit can be assumed circular, \mbox{$\pi(\ei) = \delta(\ei)$.} In practice, we sample the latter from the very narrow prior \mbox{$\pi(\ei) = U(0,0.01)$} for numerical purposes.

The system orientation angles are sampled isotropically and defined in radians throughout the rest of this paper (except where otherwise specified). The inclination follows from \mbox{$\pi(\cos\inci) = U(0, 1)$}, while the argument of periastron and the longitude of the ascending node are both drawn uniformly, \mbox{$\pi(\omegai) = \pi(\Omegai) = U(0, 2\pi)$}.

For the natal kick, we sample uniformly up to 400~\kms, \mbox{$\pi(\vkick/\kms) = U(0,400)$}, under the assumption that \acp{BH} in binaries should never attain kicks larger than this. Mass loss during \ac{CC} is treated agnostically as a fraction of the total progenitor mass, \mbox{$\pi(\fdMTwo) = U(0,1)$}, where \mbox{$\mTwof = (1-\fdMTwo)\mTwoi$}.  Note that this is related to, but distinct from, the traditional \mbox{``fallback fraction''}, and it does not depend on any assumed explosion mechanism.  In this way, we do not impose any assumptions about \ac{BH} formation, the proto-\ac{NS} mass, or fallback.  The natal kick direction is assumed isotropic in the reference frame of the exploding star, such that \mbox{$\pi(\phi) = U(0, 2\pi)$} and \mbox{$\pi(\cos\theta) = U(-1,1)$}.  The true anomaly at the moment of explosion is sampled uniformly for numerical simplicity, \mbox{$\pi(\nui) = U(0, 2\pi)$}, but samples are later reweighted to account for the time spent at each point in the orbit (see Eq.~\ref{eq:nu_dot}), with reweighting factor,
\begin{equation}
  w = \frac{(1 - \ei^2)^{3/2}}{ (1 + \ei \cos(\nui))^2}.
\end{equation}

For all azimuthal angles -- those defined on the range [0,2$\pi$) -- we sample from two normal distributions to determine a coordinate pair, and then calculate the angle in the plane for this coordinate, to avoid wrapping issues around the distribution boundaries. 
If the velocity dispersion of the birth association is included, the distributions for the velocity components are treated as priors on the sampled birth velocities \vrai, \vdeci, and \vri (see Sec.~\ref{sec:birth_velocity}).

\subsubsection{Likelihoods}
\label{sec:likelihoods}


The likelihoods for all non-angular variables are Gaussian distributions, with mean and standard deviation taken from the reported measurements.
In practice, sampling for these variables is done with a Cauchy distribution using the same arguments, and later reweighted back to a Gaussian using the likelihood ratio, since the stronger support in the tails of a Cauchy distribution allows the MCMC to better explore the low-probability regions. 

For all angular variables, the likelihoods follow a modified von Mises distribution, which is equivalent to a normal distribution wrapped around the domain [0,$2\pi$). As with the non-angular case, we sample in practice from a modified Cauchy distribution with domain [0,$2\pi$) to help with sample coverage, and then reweigh the samples to recover the von Mises distribution.
Extending the likelihoods to allow for more flexible functional forms is left to a future study.

\subsection{Degrees of freedom}
\label{sec:degrees_of_freedom}

\begin{figure*}[hbt!]
\centering
\includegraphics[width=0.8\textwidth]{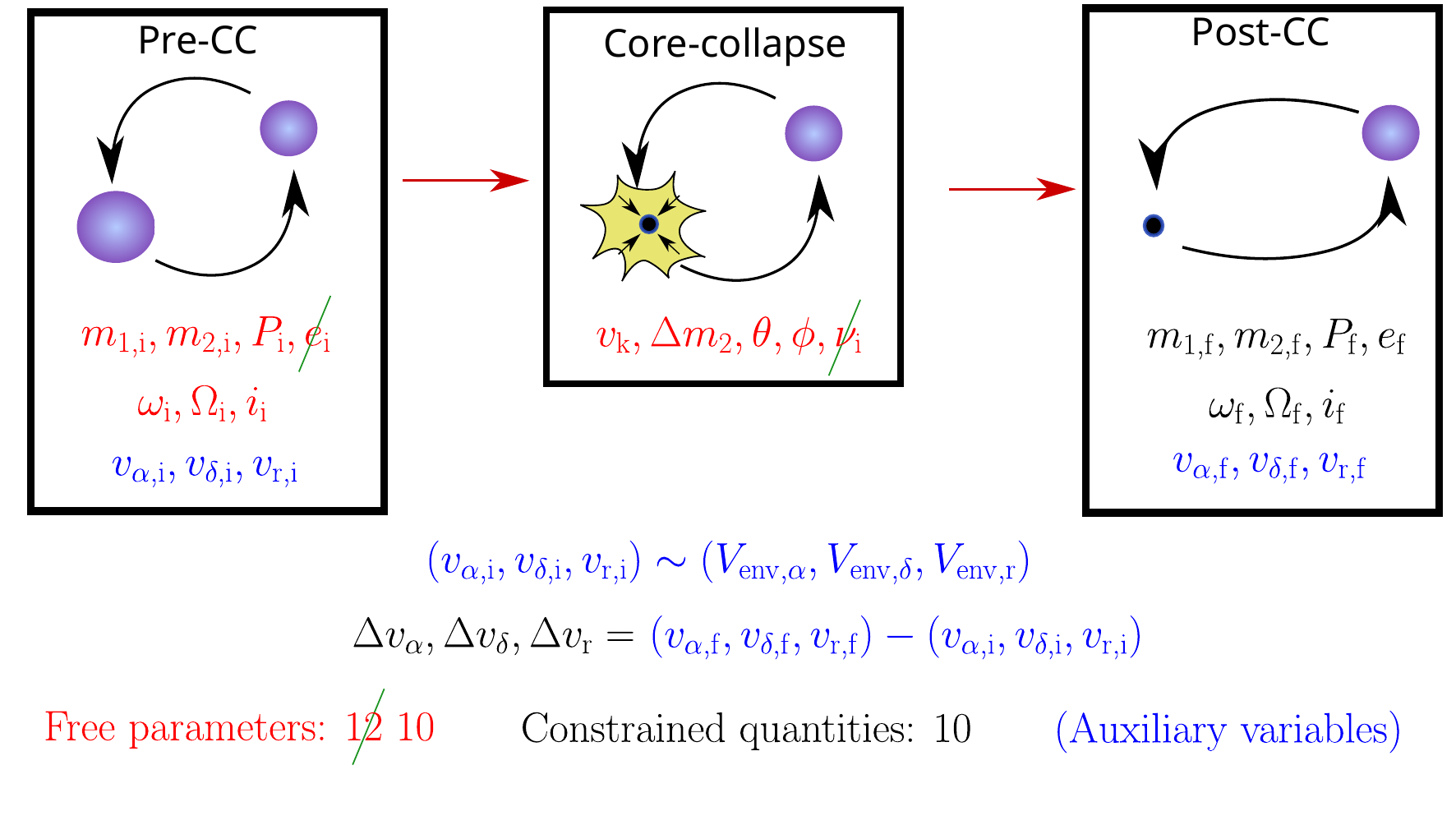}
\caption{
\textbf{A graphic showing the connection between the free parameters and the constrained quantities in this study.}
The free parameters (shown in red) describe the pre-collapse orbit and the \ac{CC} itself.
The constrained quantities (shown in black) are either directly measured or are derived from auxiliary quantities (shown in blue), in which case they need to be treated carefully.
The auxiliary post-\ac{CC} system velocities \vfVec are directly measured but unconstraining on their own.
As a proxy for the inaccessible initial velocity components \viVec, we assume that these are consistent with (and can thus be drawn from) the measured velocity dispersion of the birth association \venvVec of the binary.
The components of the net velocity vector \deltavVec are constrained quantities. 
In the best case observing scenario, we can obtain 10 distinct parameter constraints from the observations, but there are still 12 free parameters. 
If we further assume that the pre-collapse orbit was circular (which nullifies the true anomaly $\nui$), we reduce the number of free parameters to 10, and the solution can be uniquely determined. 
}
\label{fig:cartoon}
\end{figure*}

Altogether, forward modeling of the orbital elements due to the \ac{CC} involves twelve degrees of freedom, of which seven describe the pre-collapse orbit and orientation  (\Peri, \ei, \mOnei, \mTwoi, \omegai, \Omegai, \inci) and five describe the \ac{CC} itself (\dMTwo, \vkick, $\theta, \phi$, and \nui).
This setup is illustrated in Fig.~\ref{fig:cartoon}.
In the best case scenario, when a system has both spectroscopic observations and an astrometrically resolved orbit, we can obtain seven independent constraints (\Perf, \ef, \mOnef, \mTwof, \omegaf, \Omegaf, \incf). 
Assuming the proper motion and parallax have been obtained, the post-\ac{CC} systemic velocity vector is also measured, however as noted in Sec.~\ref{sec:observable_quantities}, this measurement is not useful by itself. 
If we can estimate the system's birth velocity vector from its host association, we obtain three more constraints, totaling ten altogether.

If we can further assume that the pre-collapse orbit was circular, $\ei\approx0$, then the true anomaly at the moment of collapse \nui becomes irrelevant.  More precisely, we constrain only the sum $\omegai + \nui$, but not the individual terms. In this case, we reduce the number of free parameters to ten, and the problem has as many constraints as degrees of freedom. 

In such circumstances, it is possible that the mapping between initial and final parameters is a bijection, in which case the free parameters could be estimated to arbitrarily high precision. We do not attempt to prove this, and in any event, finite measurement uncertainty is an unavoidable barrier to such arbitrarily high precision, particularly for the pre-\ac{CC} velocities which are by necessity limited by the width of the dispersion in the birth association. 
Nevertheless, we demonstrate with our injection tests that, at least for the cases under consideration, when measurement uncertainties and the birth dispersion are taken to be very narrow, we can recover the injected values precisely. 


\subsection{Estimating the system and birth velocity distributions}
\label{sec:birth_velocity}

\begin{figure}
\centering
\includegraphics[width=\columnwidth]{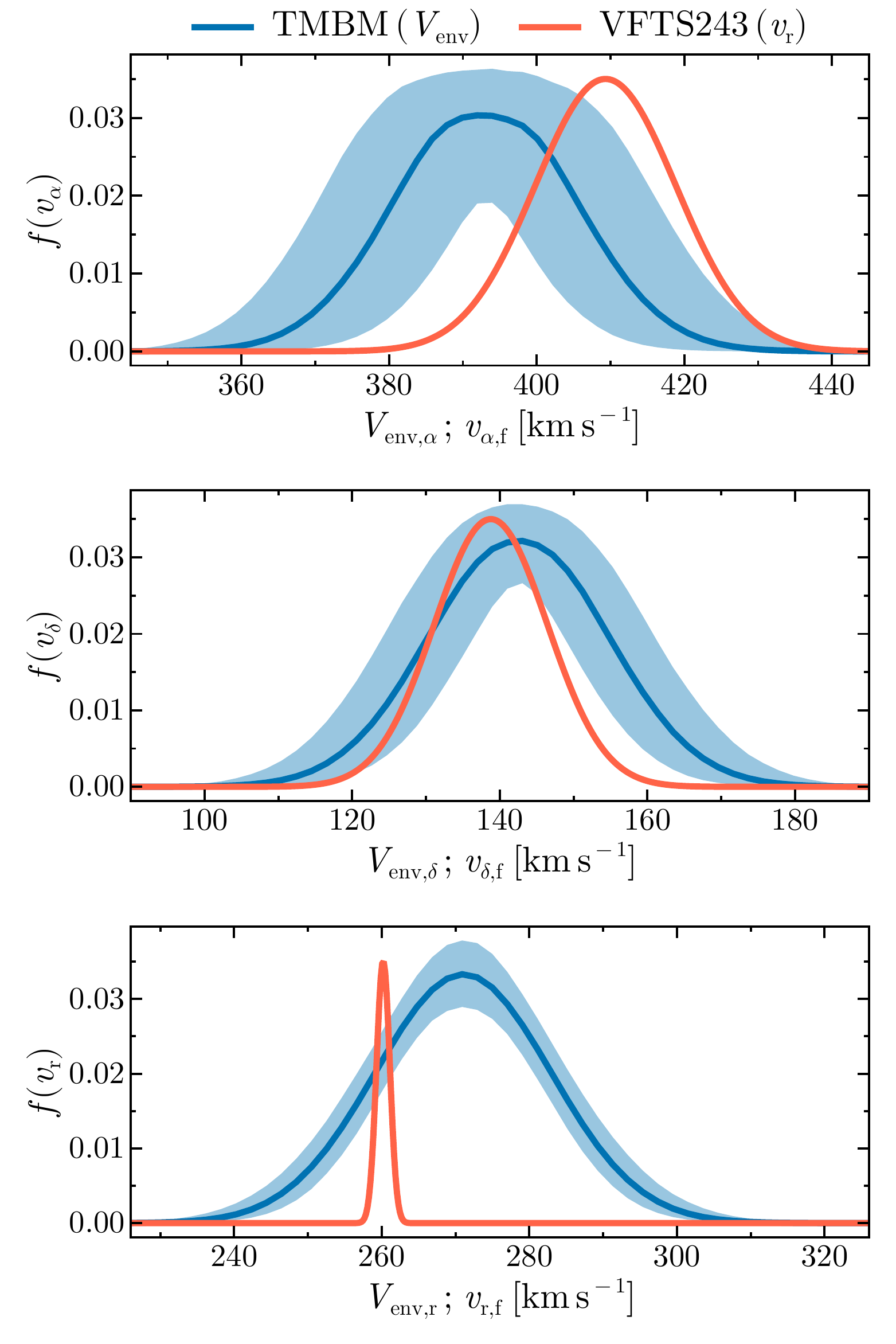}
\caption{
\textbf{Velocity distributions for VFTS 243 and its host association.}
The measured velocity components for VFTS 243 (red), compared to the inferred dispersion of its birth association, the Tarantula nebula (blue, arbitrary scaling), estimated from the massive binaries in 
the TMBM survey. 
Panels from top to bottom show the velocity 
in the direction of increasing RA, 
and the direction of increasing Dec, and in the radial direction.
For the Tarantula dispersion, a 90\% credible interval of the inferred distributions is shown.
}
\label{fig:vdists}
\end{figure}

To measure the three dimensional velocity of a binary system we make use of radial velocity, parallax, and proper motion measurements. 
Information on the natal kick cannot be inferred from the velocity of the system alone, we need to know the birth velocity of the binary to determine the change in the systemic velocity caused by the \ac{CC}. 
Here, we describe the methodology of measuring the present day velocity and estimating the birth velocity, using VFTS 243 as an illustrative example. 
VFTS 243 is extragalactic, and thus requires some care in deriving the transverse velocity solutions, as will also be the case for the systems in the \bloem sample. 

VFTS 243 was first identified in the VLT-FLAMES Tarantula survey \citep{Evans_etal.2011_VLTFLAMESTarantulaSurvey, Sana_etal.2013_VLTFLAMESTarantulaSurvey} through large variations in the measured \ac{RV} in different epochs. 
Further follow-up observations from the Tarantula Massive Binary Monitoring (TMBM) program allowed a characterization of its orbital parameters \citep{Almeida_etal.2017_TarantulaMassiveBinary}; the \ac{RV} of its barycenter was reported to be $\vrf=261.50\pm0.42~\kms$. 

For the tangential velocity components of VFTS 243, we make use of the results from the third \gaia data release (DR3, \citealt{GaiaCollaboration_etal.2023_GaiaDataRelease}).  \gaia reports correlated posterior solutions for the parallax and proper motion components for any given source. Unfortunately, since VFTS 243 is extragalactic, the reported parallax is negative, $\pax=-0.047\pm0.024~\mas$, which generally indicates that only a lower limit of $\sim10~\kpc$ can be obtained for the distance to the source \citep{Lindegren_etal.2021_GaiaEarlyData}.
The distance to the LMC is in fact $\dLMC = 49.69\pm 0.63~\kpc$, measured independently with eclipsing binaries by \citet{Pietrzynski_etal.2019_DistanceLargeMagellanic}.

It is tempting to ignore the parallax solutions completely and directly combine the proper motion data with this independent distance measurement to derive the transverse velocity for VFTS 243. 
However, since the proper motion and parallax solutions are correlated to each other, samples in the posterior space with parallax values that are too low (high) compared to the true parallax are associated to proper motions that are proportionally too high (low).
The reported proper motions, and thus the derived transverse velocities, therefore have a measurement uncertainty that is broadened by the inclusion of these extremal samples. 

Instead, we perform an MCMC where the model parameters $\vec{\theta}$ are the distance \dLMC and the two proper motions, \pmra and \pmdec. 
We use a normal distribution for the distance and broad uniform priors for the proper motions,
\begin{flalign*}
    \dLMC  ~ [\kpc  ] &\sim \curlyN(49.69, 0.63) \\
    \pmra  ~ [\masyr] &\sim U(-5,5)     \\
    \pmdec ~ [\masyr] &\sim U(-5,5).
\end{flalign*}
The likelihood on the measured parallax and proper motions is taken to be a multinormal distribution, \begin{eqnarray} \curlyL(\vec{d}|\vec{\theta})=\curlyN_3(\vec{\theta},\vec{\Sigma}),\label{equ:mulikelihood}
\end{eqnarray}
where $\vec{d}$ corresponds to the reported values of parallax and proper motions and $\vec{\Sigma}$ is the covariance matrix for these three quantities, provided by \gaia DR3. From the resulting posterior distributions, we calculate the tangential velocities of VFTS 243, $\vraf=409.3\pm9.6~\kms$ and $\vdecf=138.8\pm7.6~\kms$. In this way, we extract from the \gaia posteriors only the subset of samples which are consistent with the independently measured distance, and the resulting transverse velocity distributions are considerably more precise. 

To estimate the birth velocity, we extend this same treatment to a larger sample. The prior on the birth velocity is the velocity dispersion of the host, \venvVec, which is assumed to be isotropic and normally distributed,
\begin{eqnarray}
\viVec \sim \venvVec = \mathcal{N}_3(\muenv,  \sigmaenv^2 I).
\label{equ:vprior}
\end{eqnarray}

The velocity dispersion \venvVec is in turn calculated using a representative sample of systems within the Tarantula nebula with measured proper motions.
We again perform an MCMC to excise the regions of the \gaia posteriors with parallax values that are inconsistent with the independently measured distance.
Here, the model parameters are \dLMC, \muenv, \sigmaenv and the velocities \venvk of each system $k$ in the representative sample. 
We use the same prior for \dLMC as above.
For \muenv and \sigmaenv we use broad uniform priors, while the prior for the velocities \venvk follows from Eq.~\ref{equ:vprior} with \muenv and \sigmaenv as hyperparameters.

Some care is required in the choice of systems to include in the representative sample, as this may have a non-negligible impact on the resulting environment velocity distribution. 
For the radial velocity distribution, we choose only the SB2 binaries from the TMBM survey, reported by \citet{Almeida_etal.2017_TarantulaMassiveBinary}, as well as the ones reclassified from SB1 to SB2 using spectral disentangling by \citet{Shenar_etal.2022_TarantulaMassiveBinary}.
We do not include single stars or SB1 binaries, as these populations may be contaminated by systems that have previously been affected by their own natal kicks.

For the transverse velocities, we take the proper motion data for the same systems from \gaia DR3, but we take as additional quality cuts that the reported parallax is within 1-$\sigma$ of the LMC parallax, that both proper motions have errors smaller than 0.1~\masyr and that the Renormalized Unit Weight Error (aka RUWE) of its fit is <1.4.
For each object in the sample, we consider a normal likelihood for the \ac{RV} measurement, and for those that satisfy the quality cuts we include a likelihood on their parallax and proper motion as in Eq.~\ref{equ:mulikelihood}. 
The total likelihood is determined as the product of all of the individual likelihoods.
Because of these additional quality cuts, the number of systems included in the transverse velocity distributions is smaller than that of the radial velocity distribution.

In Fig.~\ref{fig:vdists}, we show the distributions of the three systemic velocity components \vfVec of VFTS 243 in red.  In blue, we show the inferred dispersion components \venvVec for its host, the Tarantula nebula. We can see that the velocity of VFTS 243 does not deviate significantly from the dispersion of its environment, which plays an important role in constraining its natal kick. There is a larger uncertainty in the distribution of tangential velocities of the environment, compared with the uncertainty in the \ac{RV} distribution, owing to the smaller sample size, their broad spatial distribution, and the larger measurement errors compared to \acp{RV}.

For the purpose of this work, we take the mean values of \sigmaenv and of each component of \muenv, and use these to determine normal priors for each initial velocity component when inferring kicks,
\begin{alignat*}{3}
    \vrai ~[\kms]  &\sim \venvra  & &= \curlyN(393, 12) \\
    \vdeci ~[\kms] &\sim \venvdec & &= \curlyN(143, 12) \\
    \vri ~[\kms]   &\sim \venvr   & &= \curlyN(271, 12). 
\end{alignat*}

A potential future improvement is to treat \muenv and \sigmaenv as hyperparameters, using as priors the posterior distributions that were inferred from the SB2 sample. 
The methods described here to determine the 3D velocity of VFTS 243 as well as the environment birth velocity distribution can be applied similarly to sources identified in the SMC in the \bloem survey, using a corresponding parallax measurement (e.g., \citealt{Graczyk_etal.2020_DistanceDeterminationSmall}). 

We emphasize that the choice of systems is somewhat subjective and needs to be considered on a case-by-case basis. Indeed, VFTS 243 could have been born in NGC 2060, which has a slightly higher systemic radial velocity of about 277~\kms \citep{Sana_etal.2022_VLTFLAMESTarantulaSurvey}. The SB2 population may also potentially be contaminated by dynamical runaways, artificially increasing the derived dispersion \citep{Lennon_etal.2018_GaiaDR2Reveals, Stoop_etal.2024_TwoWavesMassive}. In cases where the number of available SB2 systems is low, it may also be desirable to include single stars or SB1s despite the potential for contamination. Such considerations will be important in future analyses for constraining the birth velocities of \ibhbs.

\section{Injection tests}
\label{sec:injection_tests}

\begin{table}
\caption{
\textbf{The pre-collapse input parameters for the mock injection systems.}
All parameters are identical between the two mock systems except for the initial eccentricity. 
Horizontal lines distinguish initial parameters for the orbit, 
the pre-collapse velocities, and parameters associated to the \ac{CC} (including the true anomaly \nui at the moment of explosion). 
}
\label{tabl:input_params_test_system}
\centering
\begin{tabular}{l l l}
\hline\hline
\text{Parameter}     & \text{Value} &  \text{Description}  \\
\hline
$\mOnei$  [$\Msun$]  & 15         & Companion mass                      \\   
$\mTwoi$  [$\Msun$]  & 15         & \ac{BH} progenitor mass             \\   
$\Peri$   [day]      & 50         & Initial period                      \\   
$\ei$                & 0.0 or 0.3 & Initial eccentricity                \\  
$\Omegai$ [rad]      & 3.00       & Initial longitude of the            \\   
                     &            & \;\; ascending node                 \\  
$\omegai$ [rad]      & 6.00       & Initial argument of \\   
                     &            & periastron      \\   
$\inci$   [rad]      & 1.05       & Initial inclination                 \\   
\hline                                  
$\vrai$  [\kms]    & 200.13     & Initial system velocity \\   
$\vdeci$  [\kms]   & 100.94     & \;\;components in RA, Dec,          \\   
$\vri$  [\kms]     & 51.53      & \;\;and radial directions           \\   
\hline                                 
$\dMTwo$  [$\Msun$]  & 5          & \ac{CC} mass loss                   \\   
$\vkick$  [\kms]     & 50         & Natal kick                          \\   
$\theta$  [rad]      & 2.01       & Kick polar angle                 \\   
$\phi$    [rad]      & 0.995      & Kick azimuthal angle             \\   
$\nui$    [rad]      & 0.436      & True anomaly at \ac{CC}             \\   
\hline

\end{tabular}
\end{table}

\begin{table}
\caption{
\textbf{The post-\ac{CC} values for the mock injection systems.}
Parameters and their units are shown in column 1. Columns 2 and 3 show the values calculated for these parameters for the circular (\ei= 0.0) and eccentric (\ei= 0.3) injection tests, respectively. Column 4 shows the uncertainty, which is chosen to be illustrative, and which is relative or absolute depending on the parameter (reflecting differences in observational uncertainty estimation).
}
\label{tabl:output_params_test_system}
\centering
\begin{tabular}{l l l l}
\hline\hline
\text{Parameter}& \text{Value}& \text{Value} & \text{Uncertainty} \\
& \text{($\ei=0.0$)}& \text{($\ei=0.3$)} &  \\
\hline
$\Perf$ ~[day]       &  55.65  &  61.44 &  1\%   \\  
$\mOnef ~[\Msun$]    &  15.0   &  15.0  &  1\%   \\  
$\ef$               &  0.150  &  0.338 &  10$^{-3}$ \\
$\KOne$ ~[\kms]      &  63.20  &  62.83 &  1     \\  
$\Omegaf$ ~[rad] &  3.040  & 3.030  &  1     \\  
$\omegaf$ ~[rad] & 1.792   &  0.1943 &  1     \\  
$\incf$   ~[rad]   &  1.280  &  1.216 &  1     \\  
$\vraf$ ~[\kms]    &  190.81 &  188.1 &  1     \\  
$\vdecf$ ~[\kms]   &  75.01  &  72.66 &  1     \\  
$\vrf$ ~[\kms]     &  67.85  &  72.53 &  1     \\  
\hline
\end{tabular}
\end{table}

In the ideal scenario, observations exist for an \ibhb that can constrain all of the intrinsic and extrinsic parameters that describe its orbit, its orientation, and its velocity relative to birth. 
More commonly, some observations may not be available for a given binary, and only a subset of these data are known. 
Here, we use injection tests to explore how well we recover the parameters of interest as a function of the available observations and of assumptions about the environment and adopted prior distributions.
We performed several injection analyses using two mock binaries: one which was circular pre-collapse (\ei=~0.0, discussed below) and the other moderately eccentric (\ei=~0.3, see Appx.~\ref{appx:injection_analysis_extras}). 
The input parameters of both systems are described in Table \ref{tabl:input_params_test_system}. 

We forward model the \ac{CC} in these binaries (as described in App.~\ref{appx:explosion_math}), then add to the calculated system velocity a birth velocity drawn randomly from a specified environment dispersion distribution. 
We assume for convenience that the environment velocity distribution is described by a 3D Gaussian with mean $\muenv = (100, 200, 50)$~\kms (chosen for ease of validation), and a default dispersion $\sigmaenv = 1~\kms$ in all directions. The dispersion is small for illustrative purposes, but we explore more realistic dispersion values in Sec.~\ref{sec:velocity_dispersion_birth_host}.

We inject typical measurement uncertainties into the post-\ac{CC} properties and then perform the MCMC inference. 
The derived post-\ac{CC} properties for both systems are shown in Table \ref{tabl:output_params_test_system}, along with the adopted measurement uncertainties.

\subsection{Observational categories}
\label{sec:observational_categories}

\begin{table*}
\caption{
\textbf{Summary of models and parameters.}
The parameters that are constrained for \ibhbs depend on the available observations. We only consider binaries that are detected with spectroscopy so all model names include an ``S''. For those that have an astrometrically resolved orbit, we append an additional ``O'' to the model name.
If the host association where the binary was born can be identified, it is possible that velocity constraints can be extracted. 
If only radial velocity measurements exist for both the system and host, the model has an appended ``$\mathrm{V}_\mathrm{R}$''.
If only transverse velocity measurements exist for both the system and host, the model has an appended ``$\mathrm{V}_\mathrm{T}$''.
If both exist, the model has ``$\mathrm{V}_\mathrm{3}$'' appended.
}
\centering
\begin{tabular}{l l|| 
p{.7cm}|p{.7cm}|p{.7cm}|p{.7cm}|p{.7cm}|p{.7cm}|p{.7cm}|p{.7cm}} \hline\hline
\multirow{2}{*}{\shortstack{Category \& \\ \;\; Parameter}}
& \multirow{2}{*}{\shortstack{ \\ Symbol}}
& \multicolumn{8}{c}{Models} 
\\
\cline{3-10} & 
&  \modelS
&  \modelSVr
&  \modelSVt
&  \modelSVthree
&  \modelSO
&  \modelSOVr
&  \modelSOVt
&  \modelSOVthree
\\
\hline
Spectroscopy (S)  \\ 
\hline 
\;\;\;\;Orbital period & $\Perf$                   &*&*&*&*&*&*&*&*\\
\;\;\;\;Orbital eccentricity & $\ef$               &*&*&*&*&*&*&*&*\\
\;\;\;\;RV semi-amplitude & $\KOne$        &*&*&*&*&*&*&*&*\\
\;\;\;\;Argument of periastron & $\omegaf$         &*&*&*&*&*&*&*&*\\
\;\;\;\;Mass of star\tablefootmark{a}
& $\mOnef$           &*&*&*&*&*&*&*&*\\
\hline
Resolved Orbit (O) \\
\hline
\;\;\;\;Orbital inclination & $\incf$              & & & & &*&*&*&*\\
\;\;\;\;Longitude of the & $\Omegaf$               & & & & &*&*&*&*\\
\;\;\;\;\;\; ascending node &&&&&&&&&\\
\hline
Velocities (V)    \\ 
\hline
\;\;\;\;Radial velocity\tablefootmark{b}
& $\deltavr=\vrf - \vri$              & &*& &*& &*& &*\\
\;\;\;\;Transverse velocities\tablefootmark{b}
& $\deltavra =\vraf - \vrai$   & & &*&*& & &*&*\\
& $\deltavdec=\vdecf - \vdeci$   & & &*&*& & &*&*\\
\hline
\label{tabl:parameter_constraints}
\end{tabular}
\tablefoot{
\tablefoottext{a}{A spectroscopic SB1 can only constrain the mass function $f(\mOnef)$. Additional stellar spectral modeling is required to get an estimate of \mOnef, typically either through an estimate of $\log(g)$ or stellar track fitting.} 
\tablefoottext{b}{
The constraint \deltavr depends on the observed radial velocity \vrf and the inferred initial velocity \vri, which is assumed to be consistent with the birth environment radial velocity dispersion \venvr.
The transverse velocity constraints \deltavra and \deltavdec are treated similarly, except that they are derived from the parallax and proper motion, while the radial velocities are measured spectroscopically.}
}
\end{table*}

\begin{figure*}[htp]
  \centering
  \subfigure[
  Initially circular binary, without a resolved orbit.
  ]{\includegraphics[scale=.85]{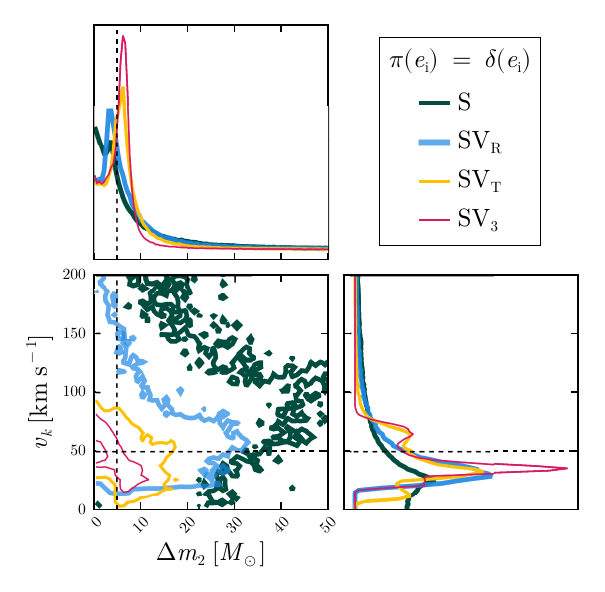}}
  \hfill
  \subfigure[
  Initially circular binary, with a resolved orbit.
  ]{\includegraphics[scale=.85]{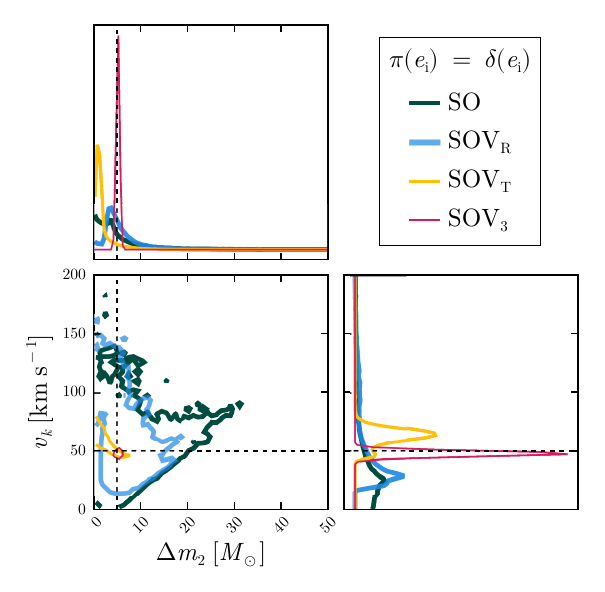}}
  \caption{
  \textbf{The impact of including different observations in a mock circular binary.}
  Posteriors on the natal kick (\vkick) and mass loss (\dMTwo), when the various observational categories are included. 
  Observational categories include spectroscopy (S), resolved astrometric orbit (O), and constraints on any of radial velocity (V$_\mathrm{R}$), transverse velocities (V$_\mathrm{T}$), or all three velocity components (V$_\mathrm{3}$).
  All categories include spectroscopy (see Sec.~\ref{sec:observational_categories} for justification).
  Categories without (with) the astrometric orbit are shown on the left (right).
  Colors correspond to differences in the available velocity constraints: 
  when there are no velocity constraints (\modelS, \modelSO: dark green),  when only the \acp{RV} are constrained (\modelSVr, \modelSOVr: blue),
  when only the transverse velocities are constrained (\modelSVt, \modelSOVt: yellow),
  and when there are measurements in all three directions (\modelSVthree, \modelSOVthree: red).
  All results here apply for the circular mock system; the results for the eccentric mock system are shown in Fig.~\ref{fig:injection_corner_plots_obs_categories_ecc}.
  Black dashed lines show the true values.
  Contours capture the 90\% confidence interval of the 2D distributions.
  }
\label{fig:injection_corner_plots_obs_categories_circ}
\end{figure*}


We consider eight models which are summarized in Table~\ref{tabl:parameter_constraints}. The models \{\modelS, \modelSVr, \modelSVt, \modelSVthree, \modelSO, \modelSOVr, \modelSOVt, \modelSOVthree{}\} are named according to the observational categories they include.  The ``S'' indicates that the binary is observed spectroscopically as an SB1. All variations include spectroscopy, as this is required in order to infer \mOnef from stellar atmosphere modeling, and thus break the degeneracy of the mass function.  In principle, the inference could be done without stellar modeling using just the mass function, however this would provide very weak constraints on the parameters of interest. The ``O'' is included when there is a resolved orbit from a close system with an astrometric binary solution. A ``V$_\mathrm{R}$'', ``V$_\mathrm{T}$'', or ``V$_\mathrm{3}$'' is appended if the net system velocity change can be constrained in the radial direction \deltavr, both of the transverse directions \deltavra and \deltavdec, or all three, respectively (we do not consider cases where only one of the transverse velocity directions is constrained). These constraints are only included if a birth association has been identified and if the indicated components are measured in both the source and the host. 

Fig.~\ref{fig:injection_corner_plots_obs_categories_circ}a shows the resulting posteriors of the models that exclude the constraints from a resolved astrometric orbit, while Fig.~\ref{fig:injection_corner_plots_obs_categories_circ}b shows models that include these constraints.
With more constraints, the posterior distributions tighten around the injected values, reducing the covered area by over an order of magnitude from the worst to best case (\modelS to \modelSOVthree).
However, there are also some biases that appear when certain observations are omitted. 
Note, for example, model \modelSOVt in Fig.~\ref{fig:injection_corner_plots_obs_categories_circ}b. 
The 90\% \ac{CI} correctly includes the true value in the 2D posterior, but the 1D distributions suggest greater support for a stronger natal kick and less mass loss.
A systematic investigation into the cause of these biases is left to a future study.

\subsection{Impact of velocity dispersion in the birth cluster}
\label{sec:velocity_dispersion_birth_host}

\begin{figure}
\centering
\includegraphics[width=\columnwidth]{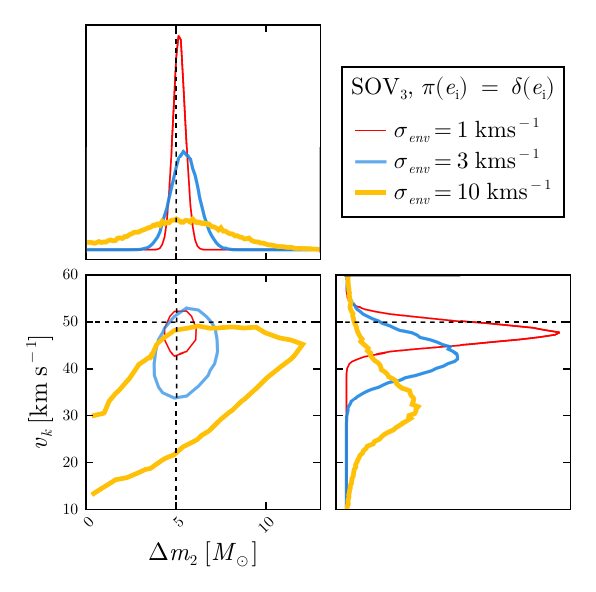}
\caption{
\textbf{Inference variations due to the birth environment velocity dispersion}.
Posteriors on the natal kick \vkick and mass loss \dMTwo, when the birth environment has velocity dispersion in all three directions equal to 1, 3, or 10 \kms (red, blue, and yellow lines, respectively).
The model shown here is for the initially circular mock system with the circular prior, in the best case scenario \modelSOVthree.
Black dashed lines show the true values.
}
\label{fig:dispersion}
\end{figure}

In Fig.~\ref{fig:dispersion}, we investigate the effect of velocity dispersion in the host association. In the default mock systems, we used a narrow velocity dispersion of 1~\kms in all three directions to help illustrate the inference capabilities.  Typical host clusters may have higher velocity dispersions, closer to 5~\kms \citep{Henault-Brunet_etal.2012_VLTFLAMESTarantulaSurvey}, while field OB stars may have velocity dispersions up to 10~\kms.  \citep{Carretero-Castrillo_etal.2023_GalacticRunawayBe}. We show the posteriors for \dMTwo and \vkick when the dispersion is allowed to vary between 1, 3, and 10~\kms.  In all cases, the circular mock system is used together with the circular prior, in the best case observing scenario (\modelSOVthree). 

As the dispersion increases, the injected values no longer lie within the 2D 90\% \ac{CI}.  In particular, the inference finds more support for low kick magnitudes, which may not be surprising given that more of the post-\ac{CC} velocity can be attributed to a larger velocity at birth.  However, we caution against drawing deep conclusions from this single test and simply argue that biases may arise from the broadening of environment dispersions. As before, a deeper investigation into the precise nature of these biases is left to future work.

\subsection{Impact of eccentricity}
\label{sec:impact_of_eccentricity}

\begin{figure*}
  \subfigure[
  Circular binary with circular prior.
  ]{\includegraphics[scale=.8]{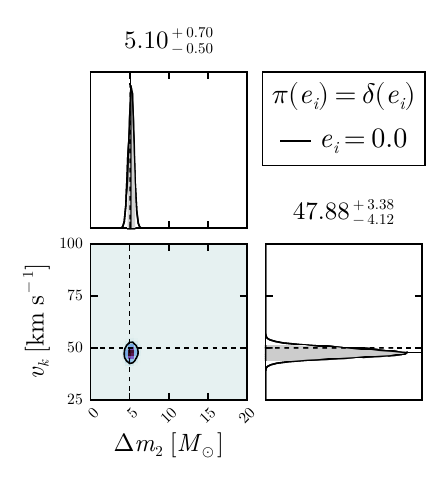}}
  \hfill
  \subfigure[
  Circular binary with eccentric prior.
  ]{\includegraphics[scale=.8]{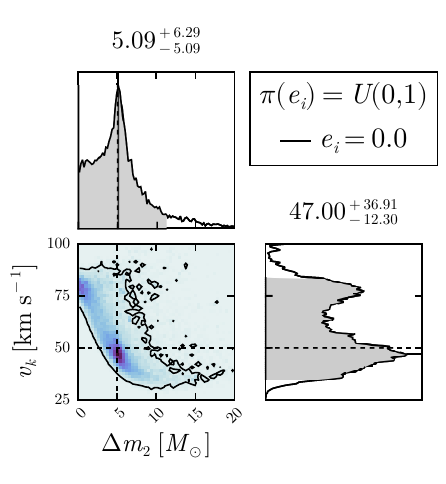}}
  \hfill
  \subfigure[
  Eccentric binary with circular prior.
  ]{\includegraphics[scale=.8]{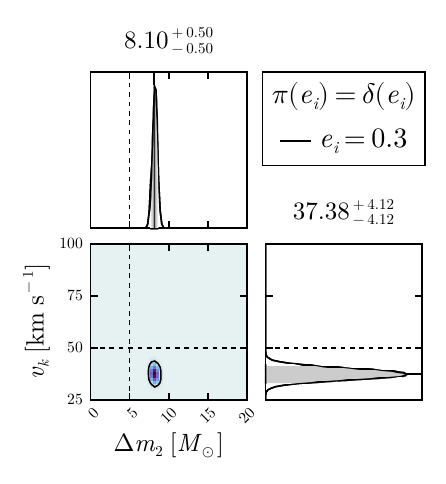}}
\caption{
\textbf{The role of eccentricity in a mock binary, in the best case observing scenario.}
Corner plots showing \dMTwo and \vkick posteriors in the best case observing scenario (\modelSOVthree, see Sec.~\ref{sec:observational_categories}).
In panel (a), we correctly assume a circular prior for the initially circular mock system, whereas in panel (b) we used the eccentric prior for the initially circular system.
In panel (c), we use a circular prior for the initially eccentric mock system, to test the impact of prior misspecification. 
The gray shaded regions and quoted intervals indicate the 90\% \ac{CI} for the 1D distributions, while black contours indicate the 90\% \ac{CI} for the two parameter correlated distributions. 
Black, dashed lines show the true values. 
}
\label{fig:eccentricity_impact}
\end{figure*}

We explore the impact of pre-collapse eccentricity and the adopted eccentricity prior used in the inference.
In Fig.~\ref{fig:eccentricity_impact}, we show several corner plots for the parameters \dMTwo and \vkick, using different eccentricity treatments. 
In all cases, we use the best case observing scenario \modelSOVthree and the default velocity dispersion of 1~\kms.

In panel (a), we assumed a circular prior for the initially circular mock system.
In this case, the number of free parameters exactly matches the number of measured observables, and we recovered all of the injected quantities with a high degree of precision, limited only by the measurement uncertainty.

In panel (b) we used the eccentric prior for the initially circular system. Here, the true value is within the 90\% \ac{CI} recovered by the inference for all parameters shown and is close to the primary peak in the posterior. However, the uncertainty intervals for the 1D distributions are roughly an order of magnitude larger than in panel (a). Additionally, there is a prominent secondary peak, suggesting that the inference may find misleading solutions even when the priors contain all of the true values.  Notably, although the posterior support is broad, we can confidently rule out the low-\vkick, low-\dMTwo region associated to direct collapse.
In panel (c), we assume a circular prior for the initially eccentric system; this is a case of model misspecification. The recovered values are reported with high precision but far away from the true values. 

The eccentricity prior therefore plays a particularly important role in the analysis.  If the agnostic (eccentric) prior is used, we can rule out some parts of the parameter space to infer coarse attributes, such as whether the \ac{BH} experienced direct collapse or not, but we cannot make precise statements about the true values of the inferred parameters.  If the circular prior is used, we can determine tight constraints on the inferred parameters, however these may be very incorrect if the pre-collapse system was not in fact circular.

\section{VFTS 243}\label{sec:vfts243}

We demonstrate our inference pipeline on the LMC spectroscopic binary VFTS 243, which is one of the first widely-accepted examples of a massive \ibhb system.  Consisting of a $\approx25\;\Msun$ O-type star orbiting a $\gtrsim9\;\Msun$ \ac{BH} in a 10.4$\;$d orbit, VFTS 243 is notable for its very low eccentricity, \mbox{$e=0.017\pm0.012$} \citep{Shenar_etal.2022_XrayQuietBlack}.  The visible companion in VFTS 243 rotates super-synchronously, indicating that tidal synchronization has not occurred since \ac{BH} formation.  Since the tidal circularization timescale is shorter than the synchronization timescale, we therefore cannot attribute the low eccentricity to tidal effects \citep{Shenar_etal.2022_XrayQuietBlack}. The orbital period, corresponding to a current separation of $\sim36\;\Rsun$, is sufficiently short that the \ac{BH} progenitor should have experienced mass transfer onto the companion. Altogether, the low eccentricity appears to be indicative of a weak natal kick and minimal mass loss during the \ac{CC} event, from a progenitor that was at least partially stripped in a binary that circularized during a prior mass transfer event \citep{ Stevance_etal.2022_VFTS243Predicted, Klencki_etal.2022_PartialenvelopeStrippingNucleartimescale,Banagiri_etal.2023_DirectStatisticalConstraints, Vigna-Gomez_etal.2024_ConstraintsNeutrinoNatal}.

We focus on VFTS 243 as it represents a good test case of what will be possible with the \bloem catalog in the SMC.
As with the SMC binaries, VFTS 243 is too distant for astrometry to resolve the orbit. 
However, \bloem data will produce spectroscopic constraints on many massive binaries in the SMC, providing source and host \acp{RV}, similarly to the VLT-FLAMES survey for the Tarantula nebula. And existing \gaia astrometric data combined with an independent distance measurement provides transverse velocities for sources in both Magellanic clouds.
VFTS 243 and the upcoming BLOeM systems therefore fall into the observational category \modelSVthree introduced in Table~\ref{tabl:parameter_constraints}.
The spectroscopically-obtained parameters reported for VFTS 243 in \citet{Shenar_etal.2022_XrayQuietBlack} are reproduced in Table \ref{tabl:vfts243_obs}, together with the estimated systemic transverse and radial velocities from \citet{GaiaCollaboration_etal.2023_GaiaDataRelease} and \citet{Almeida_etal.2017_TarantulaMassiveBinary}, respectively.
For the environment velocity estimate, we follow the routine outlined in Sec.~\ref{sec:birth_velocity}.

\begin{table}
\caption{
\textbf{Properties of VFTS 243.}
Published values for the parameters of VFTS 243 and its system velocity components. Uncertainties are reported at the 1-$\sigma$ level. 
}
\label{tabl:vfts243_obs}
\centering
\begin{tabular}{ l l l l}
\hline\hline
\text{Parameter}& Value & Units & Ref. \\
\hline
$\Perf  $ & 10.40 $\pm$ 0.01  & day       & 1\\
$\ef    $ & 0.017 $\pm$ 0.012 &           & 1\\
$\mOnef\tablefootmark{a} $ & 25    $\pm$ 12    & $\Msun$   & 1\\
$\KOne  $ & 81.4  $\pm$ 1.3   & $\kms$    & 1\\
$\omegaf$ & 1.15  $\pm$ 0.925 & rad       & 1\\
\hline                                                
$\vraf  $ & 409.3 $\pm$ 9.3   & $\kms$    & 2\\
$\vdecf $ & 138.8 $\pm$ 7.6   & $\kms$    & 2\\
$\vrf   $ & 261.5 $\pm$ 0.4   & $\kms$    & 3\\
\hline
\end{tabular}
\tablefoot{\tablefoottext{a}{
We adopt for \mOnef the average between the evolutionary and spectroscopic mass estimates (see \citet{Shenar_etal.2022_XrayQuietBlack}).
}}
\tablebib{
(1)~\citep{Shenar_etal.2022_XrayQuietBlack};
(2) \citep{GaiaCollaboration_etal.2023_GaiaDataRelease};
(3) \citep{Almeida_etal.2017_TarantulaMassiveBinary}
}
\end{table}

\begin{table}
\caption{
\textbf{Priors used in the analysis of VFTS 243}. These are the default distributions used in all variations, except where noted.
They are chosen to be wide and uninformative. 
\\\hspace{\textwidth}
}
\label{tabl:vfts243_priors}
\centering
\begin{tabular}{ l l }
\hline\hline
\text{Parameter}& \text{Prior} \\
\hline
$\log(\mOnei/\Msun)$   & $U$(0.1, 3)\\
$\log(\mTwoi/\Msun)$   & $U$(0.1, 3)\\
$\log(P_i/\mathrm{d})$ & $U$(-1, 3) \\
$\ei$                  & $\delta(\ei)$ \tablefootmark{a} \\
$\cos\inci$            & $U$(0, 1)   \\
$\omegai$ ~[rad]        & $U(0, 2\pi)$ \\ 
$\Omegai$ ~[rad]        & $U(0, 2\pi)$ \\ 
\hline
$\vkick ~[\kms]$        & $U(0, 400)$ \\
$\fdMTwo$              & $U$(0, 1)    \\
$\cos\theta$           & $U$(-1, 1)    \\
$\phi$ ~[rad]           & $U(0, 2\pi)$   \\
$\nui$ ~[rad]           & $U(0, 2\pi)$\tablefootmark{b} \\ 
\hline
$\venvra  ~[\kms]$      & $\curlyN(393, 42)^{(1)}$ \\
$\venvdec ~[\kms]$      & $\curlyN(146, 40)^{(1)}$ \\
$\venvr   ~[\kms]$      & $\curlyN(270, 11)^{(2)}$  \\
\hline
\end{tabular}
\tablefoot{
\tablefoottext{a}{
The circular orbit prior is 
replaced with the eccentric prior $U$(0, 1) in the final model variation.
}
\tablefoottext{b}{The true anomaly prior is sampled uniformly for simplicity, then later reweighted so that \nui is more correctly uniform in time.}
}
\tablebib{
(1)~\citet{GaiaCollaboration_etal.2023_GaiaDataRelease}; (2) \citet{Almeida_etal.2017_TarantulaMassiveBinary}
}
\end{table}

\subsection{Inference outcomes}

\begin{figure}
\centering
\includegraphics[width=\columnwidth]{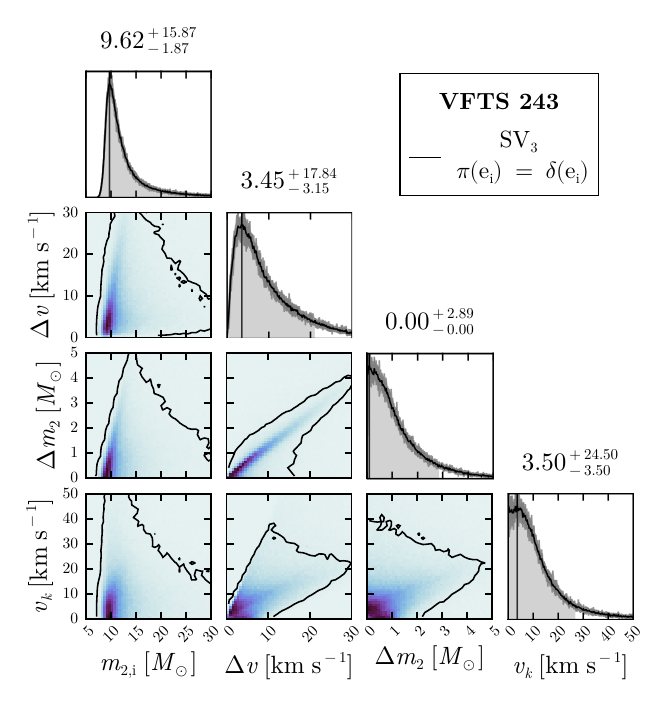}
\caption{
\textbf{The best case inference for VFTS 243}. Here, we use circular orbit prior with the \modelSVthree model, which is the best case for an extra-Galactic system, combining spectroscopy as well as the full velocity vector of the system and the dispersion of its birth environment. 
}
\label{fig:vfts243_derivables}
\end{figure}

The results of the inference for VFTS 243 are shown in Fig.~\ref{fig:vfts243_derivables}, using the \modelSVthree model and assuming a circular orbit prior.  We highlight only a few key parameters of interest, the \ac{BH} progenitor mass \mTwoi, the systemic velocity relative to the pre-collapse velocity \deltav, and the \ac{CC} mass loss \dMTwo and natal kick \vkick (posteriors for other parameters are shown in Appx.~\ref{appx:vfts243_cornerplots}). Even in the absence of a resolved orbit, the posteriors are informative and unimodal. 
As expected, the inference indicates that VFTS 243 formed with little mass loss and a weak natal kick.

\begin{figure}
\centering
\includegraphics[width=\columnwidth]{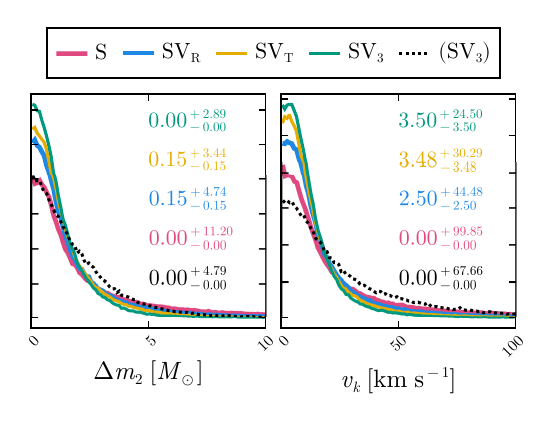}
\caption{
\textbf{Variations on the inference of VFTS 243 due to different observations and priors.}
We compare the 1D posteriors of the mass loss \dMTwo and natal kick \vkick, for VFTS 243.
We include only models which which do not require a resolved orbit (see Sec.~\ref{sec:observational_categories}).
Models \modelS, \modelSVr, \modelSVt, and \modelSVthree (solid lines) use the circular orbit prior, $\pi(\ei) = \delta(\ei)$.
We additionally include the best case observing model \modelSVthree with the full eccentricity prior, $\pi(\ei) = U(0,1)$ (black, dotted), identified as (\modelSVthree).
}
\label{fig:vfts_multihist}
\end{figure}

For completeness, we also explore models \modelS, \modelSVr, and \modelSVt to see the impact if we ignore some of the velocity components and to better compare to previous studies.
In Fig.~\ref{fig:vfts_multihist}, we show the impact of these variations together with the best case model \modelSVthree on the 1D posteriors for \dMTwo and \vkick. These models all use the circular orbit prior. We additionally show model \modelSVthree using the agnostic eccentricity prior, as a check against model misspecification in case the pre-collapse orbit was not circular. All variations are consistent with low mass loss and small natal kicks, although the 90\% credible intervals differ in some cases by a factor of four.  This highlights the straightforward interpretation of circular \ibhbs which have not been affected by tides.

\subsection{Comparison to previous studies}

Several recent studies also reported constraints on the natal kick and mass loss properties of VFTS 243 \citep{Stevance_etal.2022_VFTS243Predicted, Banagiri_etal.2023_DirectStatisticalConstraints, Vigna-Gomez_etal.2024_ConstraintsNeutrinoNatal}. 
We qualitatively confirm the results in these previous studies that this system likely formed with a relatively small amount of mass loss and a modest natal kick that is consistent with 0. 

\citet{Stevance_etal.2022_VFTS243Predicted} used pre-computed models in the BPASS population synthesis code \citep{Eldridge_etal.2008_EffectMassiveBinaries,Stanway_Eldridge.2018_ReevaluatingOldStellar} to infer a natal kick $\vkick < 33~\kms$ at 90\% confidence.
This is only marginally higher than the 26.5~\kms upper limit reported here.
Their best fit models also indicate mass loss $\dMTwo \approx 1~\Msun$, although this is not reported with an uncertainty interval.
This value is slightly larger than, but consistent with, the median value reported here.

\citet{Banagiri_etal.2023_DirectStatisticalConstraints} and \citet{Vigna-Gomez_etal.2024_ConstraintsNeutrinoNatal} both perform their inference directly on the observed properties of VFTS 243, with an agnostic treatment of the stellar and binary evolution history. 
Both include constraints on the \ac{RV} for VFTS 243, but not the transverse velocity data, thus these studies are best compared to the \modelSVr curves. 

\citet{Banagiri_etal.2023_DirectStatisticalConstraints} report a 90\% upper limit for the natal kick of either $\vkick < 72~\kms$ or $< 97~\kms$, depending on the choice of prior, which is a factor of $\sim3-4$ larger than found here.
They also report a 90\% interval in the mass loss of either $\dMTwo/\Msun \in [0.4,14.2]$ or $[0.3,14.4]$, again depending on the choice of prior, although elsewhere in the paper they claim a 90\% lower mass loss constraint of either 0.69~\Msun or 0.77~\Msun. 
These upper limits are also a factor of $\sim3-4$ higher than the upper limits reported here, and moreover our support at \dMTwo = 0~\Msun is in tension with their conclusions. 
However, this is not a direct one-to-one comparison, because \citet{Banagiri_etal.2023_DirectStatisticalConstraints} do not account for the correlation between the \ac{RV} constraint and the other parameters.

Meanwhile, \citet{Vigna-Gomez_etal.2024_ConstraintsNeutrinoNatal} report a natal kick which peaks at 0~\kms with a 68\% upper limit of 10~\kms and an ejecta mass 68\% upper limit of 1~\Msun, also peaking at 0~\Msun. 
To facilitate the direct comparison, we report here also our 68\% confidence intervals on these parameters.
At 68\% confidence, we find the natal kick $\vkick \leq 13.47~\kms$ and the mass loss $\dMTwo \leq 1.34~\Msun$, both still peaking at 0, and only slightly larger than the values found in \citet{Vigna-Gomez_etal.2024_ConstraintsNeutrinoNatal}.
However, \citet{Vigna-Gomez_etal.2024_ConstraintsNeutrinoNatal} used top-hat likelihoods in place of Gaussians on all observed parameters, which is overly restrictive, in addition to using the \ac{RV} measurement to inform the likelihood on the full systemic velocity.
Thus we consider the analysis framework here to be a more robust improvement over the previous methodology.

\section{Discussion}\label{sec:discussion}


\subsection{Importance of the pre-collapse eccentricity prior}
\label{sec:importance_preCC_eccentricity}

In our injection tests we considered two systems, one which was circular pre-collapse and one which was eccentric. 
We also considered priors on the orbital eccentricity that were either circular, $\pi(\ei) = \delta(\ei)$, or were broad and uninformative, $\pi(\ei) = U(0,1)$.  In the latter case, the posteriors become so broad that only coarse, qualitative statements can be made, even in the best case observing scenario model \modelSOVthree (see App.~\ref{appx:injection_analysis_extras}). 

However, we cannot assume that all observed \ibhbs were circular prior to the \ac{CC}.  Some may not have interacted prior to \ac{CC}, and so could have retained a natal eccentricity from the birth of the binary. Others may have been part of triple (or higher multiplicity) systems, as is expected for many massive stars \citep{Moe_DiStefano.2017_MindYourPs}.  This would have a non-trivial effect on the orbits both pre- and post-\ac{CC}, which we do not attempt to account for here. Additionally, even in the case of binaries which interacted prior to collapse, recent evidence suggests that mass transfer does not circularize binaries in all cases \citep[e.g,][and references therein]{Vigna-Gomez_etal.2020_CommonEnvelopeEpisodes}, as has been the ubiquitous assumption in binary evolution models for many decades.

There is some observational evidence for eccentric, post-interaction binaries, particularly where one component is a \ac{WR} star, such as the WR + O  binary $\gamma^2$ Vel (aka WR\,11) with a period $P=79$~d and an eccentricity $e\approx0.33$ \citep{Niemela_Sahade.1980_OrbitalElementsGam2, Eldridge.2009_NewageDeterminationG2, Lamberts_etal.2017_NumericalSimulationsInfrared}. Other such systems  have eccentricities in the range $e\in[0.1,0.4]$, and periods between 3 and 5000 days, although the post-interaction nature has been debated for some of these \citep{Marchenko_etal.1994_WolfRayetBinaryV444, Schmutz_Koenigsberger.2019_LongUninterruptedPhotometric, Richardson_etal.2021_FirstDynamicalMass, Richardson_etal.2024_VisualOrbitsWolf, Holdsworth_etal.2024_VisualOrbitsWolf}. Additionally, WR 140 is extremely eccentric, with $e=0.89$, although it is in a wide, 7.93~yr orbit which again challenges the post-interaction hypothesis \citep{Lau_etal.2022_NestedDustShells, Holdsworth_etal.2024_VisualOrbitsWolf}. Recently, \citet{Pauli_etal.2022_EarliestOtypeEclipsing} reported the discovery of AzV 476, an O4+O9.5 binary in the SMC, with $P=9.4$~d and $e=0.24$. The authors argued that AzV 476 must be post-interaction based on the similarity in component masses, which is in tension with the large difference in luminosities.

Moreover, recent theoretical work has argued for eccentricity retention during mass transfer in long period, roughly equal mass binaries \citep{Rocha_etal.2024_MassTransferEccentric}, as well as eccentricity pumping due to phase dependent Roche-lobe overflow  \citep{Sepinsky_etal.2007_InteractingBinariesEccentric, Sepinsky_etal.2009_InteractingBinariesEccentric, Sepinsky_etal.2010_InteractingBinariesEccentric} or in post-mass transfer binaries due to the presence of a circumbinary disk \citep{Vos_etal.2015_TestingEccentricityPumping, Valli_etal.2024_LongtermEvolutionBinarya}.
Given this, we might expect the true population of pre-collapse binaries to be circularized due to mass transfer up to a limiting period, above which systems can retain more eccentricity \citep{Rocha_etal.2024_MassTransferEccentric}. 
In general, we suggest caution when considering the role of pre-collapse eccentricity on the inference results from \sidekicks{}.

\subsection{Ideal candidates for observation}
\label{sec:ideal_candidates_observation}

As the population of \ibhbs grows, it will become crucial to determine which systems might provide useful constraints on \ac{BH} formation. 
A critical assumption to the inference here is that the binary orbit has not changed substantially since formation of the \ac{BH}.
Close binaries that have experienced mass transfer after \ac{CC}, or which have circularized under the influence of tidal interactions, modify key elements of the orbit and thus invalidate this assumption \citep{Shenar_etal.2022_XrayQuietBlack}.
Similarly, binaries that have traveled significantly through their galactic environments may have systemic velocities that have been modified, or potentially experienced perturbative fly-by interactions from a third body \citep{Stegmann_etal.2024_CloseEncountersWidea}.
This is unlikely to be the case for binaries with an O/B-type primary, which provide an upper age limit of at most $\sim10~\text{Myr}$, however it may be problematic for binaries with low-mass primaries, such as the three \gaia \acp{BH}. 

\gaia \ac{DR4} is expected to provide solutions to many more massive binaries, including potentially hundreds of O/B+\ac{BH} binary candidates, many of which should be useful for this type of inference  \citep{Janssens_etal.2022_UncoveringAstrometricBlack}. Such systems would require spectroscopic follow up to confirm the \ac{BH} companion, such as from the Sloan Digital Sky Survey \citep{Almeida_etal.2023_EighteenthDataRelease}. Those for which the orbit can be resolved may provide tighter constraints on the \ac{BH} formation than \ibhbs from the more distant \bloem sample. However, they may also be biased towards longer periods, where the assumption of pre-collapse circularity might not hold.  Crucially, \acp{BH} observed in any kind of binary are necessarily biased towards low, non-disruptive kicks, so that even with a large population of \ibhbs analyzed with \sidekicks{}, we will not be able to truly probe the full \ac{BH} natal kick landscape. 

While not the focus of this study, inert \ac{NS}-binaries can also be used with this inference pipeline, assuming the same conditions as described above for \acp{BH}. This would provide valuable constraints on the weak end of the \ac{NS} natal kick distribution, which is important in the formation of \ac{NS}-\acp{XRB}, binary \acp{NS}, short gamma-ray bursts, kilonovae, and other related phenomena, and complements the well-studied constraints on the stronger end of the \ac{NS} natal kick distribution.  Notably, the lower masses of \acp{NS} compared to \acp{BH} make it more challenging to distinguish \acp{NS} from low mass non-degenerate stars when interpreting the companions in SB1 binaries.

\subsection{Model dependency of the primary mass}
\label{sec:model_dependency_primary_mass}

A strength of our methodology is the lack of model dependent quantities involved. 
The priors are mostly chosen to be broad and non-informative, the \ac{CC} mechanism is treated agnostically, and the effect of the \ac{CC} on the binary is a straightforward adaptation of the 2-body problem.
However, as noted in Sec.~\ref{sec:observable_quantities}, we cannot determine the component masses in a model-independent way; we are limited by the degeneracies of the mass function.

Primary mass estimates must come from some other model dependent technique, such as the spectroscopic mass derived from the surface gravity and radius, or the evolutionary mass approximated from matching the position of the star in the HR-diagram to precomputed single star evolution tracks (see \citealt{Shenar_etal.2022_XrayQuietBlack}).  The evolutionary mass in particular may be unreliable, especially given that many of the \ibhbs are expected to have interacted prior to the formation of the \ac{BH}, in which case the structure and interpretation of the primary star may be biased \citep{Renzo_etal.2023_RejuvenatedAccretorsHave}. These systematic uncertainties propagate through the inference, and may complicate efforts to estimate \ac{CC} and pre-collapse parameters.


\subsection{Extensions to the analysis}
\label{sec:extensions_to_the_analysis}

The inference pipeline described in this study could be extended in a number of ways, to incorporate other types of binaries or observations than those studied here. 
High-mass \acp{XRB}, such as Cyg X-1, typically circularize due to tidal forces, modifying their periods and eccentricities from their values following \ac{CC}. 
However, the tidal circularization mechanism conserves angular momentum.
Thus, by treating the circularization period,
\begin{equation}
\PerCirc = \Perf(1-\ef^2)^{3/2},
\end{equation}
of the binary as an observable, instead of the period and eccentricity independently, we can perform similar inference on the high-mass \ac{XRB} population, assuming tidal forces are the dominating perturbation in the post-\ac{CC} evolution of the binary. For Cyg X-1 in particular, recent observations that resulted in a $\sim50\%$ increase in the estimated \ac{BH} mass \citep{Miller-Jones_etal.2021_CygnusX1Contains} warrant an updated investigation of the natal kick constraints, which we plan to address in an upcoming publication.

Additionally, eclipsing binaries may experience self-lensing and brighten during the eclipse, providing a tight constraint on the inclination of the binary, and possibly also an independent estimate on the \ac{BH} mass if the parallax and companion effective temperature are also constrained \citep{Masuda_Hotokezaka.2019_ProspectsFindingDetached, Chawla_etal.2024_DetectingDetachedBlack}.  Moreover, \ibhbs may be identifiable via ellipsoidal variability or relativistic beaming, constraining functions of the masses, inclination, and semi-major axis \citep{Chawla_etal.2024_DetectingDetachedBlack}. Furthermore, \citet{Mahy_etal.2022_IdentifyingQuiescentCompact} showed that for spectroscopic binaries, a large measured $v\sin i$ constrains the inclination angle to be below a given threshold to prevent the luminous star from spinning above critical rotation.  However, given that the orbital angular momentum vector may well have rotated during \ac{CC}, this is a constraint on the maximum pre-collapse inclination \inci.

\subsection{Impact on primary}
\label{sec:impact_on_primary}

Key quantities that we ignored throughout include those related to the impact of a \ac{SN} explosion on the primary star, both in the form of a mass differential \dMOne (due to ablation or accretion) or an impact velocity injection \vimp \citep{Wheeler_etal.1975_SupernovaeBinarySystems, Pfahl_etal.2002_ComprehensiveStudyNeutron,Liu_etal.2015_InteractionCorecollapseSupernova, Hirai_etal.2018_ComprehensiveStudyEjectacompanion, Ogata_etal.2021_ObservabilityInflatedCompanion}.
These may be negligible in all but the closest binary systems, though such effects could potentially be observable in late time \ac{SN} follow up observations \citep{Hirai_Podsiadlowski.2022_NeutronStarsColliding}.
If included, these would increase the size of the parameter space and broaden the posteriors on the parameters of interest.

\section{Summary and Conclusions}
\label{sec:conclusions}

Black hole natal kicks persist as an astrophysical uncertainty with broad consequences across different fields, with impacts on (among others) \ac{BH} X-ray binaries, \acp{BH} in globular clusters, free-floating \acp{BH} which could contribute to microlensing observations and dark matter estimates \citep{Cirelli_etal.2024_DarkMatter}, and binary \acp{BH} and \ac{BH}-\acp{NS} which may be detectable as gravitational-wave sources.

In this work, we presented a Bayesian inference and MCMC sampling package, \sidekicks{}, that uses a broad collection of observables for \ibhbs to constrain \ac{BH} natal kicks and mass loss, as well as the pre-collapse orbital parameters that provide useful constraints on binary evolution models. While previous works have studied the impact of a core-collapse event in a binary with an eccentric orbit \citep{Pfahl_etal.2002_ComprehensiveStudyNeutron, Hurley_etal.2002_EvolutionBinaryStars}, we expand on these efforts by tracking all of the Keplerian orbital elements of the binary together with the post-collapse velocity vector.

We argue that the inclusion of these extra observables provides critical constraints on the parameters of core-collapse that have been missed in previous analyses which only account for the post-\ac{CC} speed. In the special case when the binary was circular prior to \ac{BH} formation, and when the current orbit can be completely characterized, we demonstrate that all of the collapse and pre-collapse parameters, in particular the natal kick and mass loss, can be recovered precisely. We further explore the consequences of excluding some of these observed parameters, either through lack of observational data or physical unattainability, as well as incorrectly assuming that the pre-collapse orbit was circular. 

Finally, we demonstrate the inference capabilities of \sidekicks by inferring the properties of VFTS 243, an \ibhb in the LMC. We highlight the improvements of our method with respect to previous studies that did not correctly treat all of the orientation and velocity elements.  We report a mass loss $\dMTwo/\Msun = 0.07^{+2.82}_{-0.07}$ and a natal kick $\vkick/\kms = 0.00^{+26.45}_{-0.00}$, which is qualitatively in agreement with the previous literature values for this system.

A particular strength of the method presented in this work is that it is almost completely independent of uncertain astrophysical models, such as the details of the core-collapse mechanism and dependence on the progenitor structure, although it does depend on uncertain mass estimates for the visible star from either spectroscopic mass estimates or evolutionary models. This analysis can be readily applied to any binaries containing a massive star and a compact object, including \acp{NS}, in wide binaries with negligible tidal interactions.
Such binaries are expected to be observed in large numbers with the \bloem spectroscopic survey, as well as with the upcoming \gaia \ac{DR4} release.


\section*{Software}
Data analysis was performed with a custom inference package, \hyperlink{https://github.com/orlox/SideKicks.jl}{\sidekicks{}}, using the specific version stored at  \url{https://doi.org/10.5281/zenodo.15147091}.  
The velocity inference performed in Sec.~\ref{sec:birth_velocity} can be reproduced using the data and scripts stored at \url{https://doi.org/10.5281/zenodo.15196531},
while those for the injection tests in Sec.~\ref{sec:injection_tests} are stored at  \url{https://doi.org/10.5281/zenodo.15267344}.
\sidekicks{} made extensive use of the Bayesian inference library \textsc{Turing.jl} \citep{Ge_etal.2018_TuringLanguageFlexible} and the plotting package \textsc{Makie.jl} \citep{Danisch_Krumbiegel.2021_MakiejlFlexibleHighperformance}, all of which are written in the \juliaCode{} language \citep{Bezanson_etal.2014_JuliaFreshApproach}.
This work has made use of data from the European Space Agency (ESA) mission \gaia (\url{https://www.cosmos.esa.int/gaia}), processed by the \gaia Data Processing and Analysis Consortium (DPAC, \url{https://www.cosmos.esa.int/web/gaia/dpac/consortium}). Funding for the DPAC has been provided by national institutions, in particular the institutions participating in the \gaia Multilateral Agreement.

\begin{acknowledgements}

RW acknowledges support from the KU Leuven Research Council through grant iBOF/21/084. PM acknowledges support from the FWO senior fellowship number 12ZY523N and the European Research Council (ERC) under the European Union's Horizon 2020 research and innovation programme (grant agreement 101165213/Star-Grasp). HS and KD acknowledge the financial support from the Flemish Government under the long-term structural Methusalem grant METH/24/012 at KU Leuven. ME acknowledges funding from the FWO research grants G099720N and G0B3823N. LM acknowledges the Belgian Science Policy Office (BELSPO) for the financial support. SJ acknowledges support from International postdoctoral research fellowships of the Japan Society for the Promotion of Science (JSPS). LRP acknowledges support by grants PID2019-105552RB-C41 and PID2022-137779OB-C41 and PID2022-140483NB-C22 funded by MCIN/AEI/10.13039/501100011033 by ``ERDF A way of making Europe''. DP acknowledges financial support from the FWO in the form of a junior postdoctoral fellowship grant number No. 1256225N. AACS is supported by the Deutsche Forschungsgemeinschaft (DFG - German Research Foundation) in the form of an Emmy Noether Research Group -- Project-ID 445674056 (SA4064/1-1, PI Sander). TS acknowledges support from the Israel Science Foundation (ISF) under grant number 0603225041 and from the European Research Council (ERC) under the European Union's Horizon 2020 research and innovation programme (grant agreement 101164755/METAL).

\end{acknowledgements}

%
%
\bibliographystyle{aa}
\bibliography{bib}

\begin{appendix}

\section{Post explosion orbital elements of an eccentric binary}\label{appx:explosion_math}
\subsection{Pre-explosion properties}

\begin{figure}
\centering
\includegraphics[width=0.7\columnwidth]{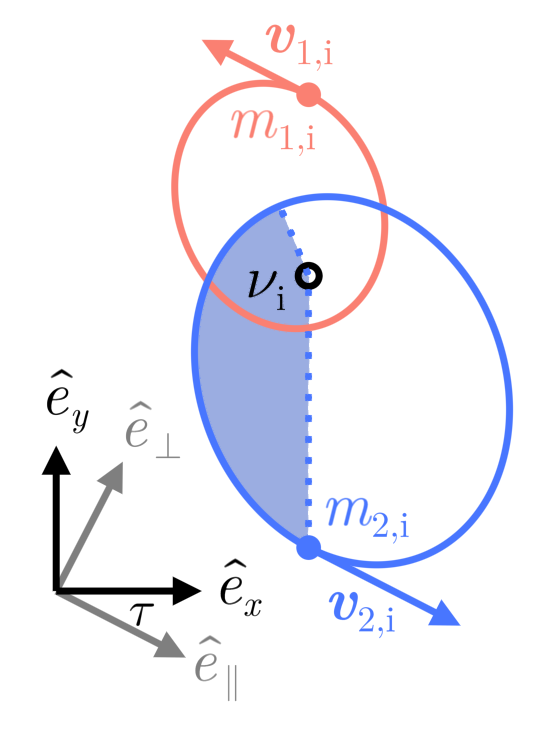}
  \caption{\textbf{Definitions of unit axes used to determine the post-explosion parameters of an eccentric binary.}
  Shown is the binary at the moment of explosion of star 2 with a true anomaly $\nui$. 
  The black circle indicates the center of mass of the system. 
  The unit axes $\ehatx$ and $\ehaty$ are defined in terms of the line joining both stars at the moment of explosion, while the unit axes $\ehatPar$ and $\ehatPrp$ are defined in terms of the direction of motion of star 2 at the moment of explosion. 
  These sets of unit vectors are related by a rotation with an angle $\tau$.
  }
     \label{appx:orbitaxes}
\end{figure}

Let us consider a binary system composed of two masses, \mOnei and \mTwoi, with a semimajor axis $\ai$ and an eccentricity $\ei$. 
We will consider the case where star $2$ undergoes an explosion at a point in the orbit with a true anomaly $\nui$. 
We will refer to properties right before and right after the explosion with subindeces `i' and `f' respectively. 
We define a coordinate frame oriented such that $\ehaty$ is a unit vector pointing from star 2 towards its companion at the moment of explosion. 
The unit vector $\ehatz$ points in the direction of the orbital angular momentum and as usual $\ehatx=\ehaty\times\ehatz$ (see Fig.~\ref{appx:orbitaxes}). 
We can then define the position vectors of both stars as a function of arbitrary true anomaly $\nu$ as
\begin{eqnarray}
\begin{aligned}
    \vec{r}_1 &=&  r_1 [-\sin(\nu-\nui)\ehatx + \cos(\nu-\nui)\ehaty] \\
    \vec{r}_2 &=&  r_2 [\sin(\nu-\nui)\ehatx - \cos(\nu-\nui)\ehaty].
\end{aligned}
\end{eqnarray}
The distance of each star to the center of mass is given by,  
\begin{equation}
    r_1 = \ai\frac{\mTwoi}{\Mi}f(\nu, \ei),\quad r_2=\ai\frac{\mOnei}{\Mi}f(\nu, \ei),
\end{equation}
where $\Mi=\mOnei+\mTwoi$ is the total mass of the pre-explosion system. The quantity $f(\nu, \ei)$ determines the instantaneous separation $\ai f(\nu, \ei)$ of the binary,
\begin{equation}
\label{eq:fnu}
    f(\nu, \ei) = \frac{1-\ei^2}{1+\ei\cos\nu}.
\end{equation}
The time derivative of the true anomaly can be derived from Kepler's second and third law, resulting in
\begin{equation}
\label{eq:nu_dot}
    \dot{\nu} = \sqrt{\frac{G\Mi}{\ai^3}}\frac{(1+\ei\cos\nu)^2}{(1-\ei^2)^{3/2}}.
\end{equation}

Combining these results, we can write the individual velocities of each component with respect to the center of mass of the binary, at the moment of explosion ($\nu = \nui$),
\begin{eqnarray}
\begin{aligned}
    \vOnei &= -\frac{\mTwoi}{\Mi}\sqrt{\frac{G\Mi}{\ai(1-\ei^2)}}[(1+\ei\cos\nui)\ehatx-\ei\sin\nui\ehaty],\\
    \vTwoi &= \frac{\mOnei}{\Mi}\sqrt{\frac{G\Mi}{\ai(1-\ei^2)}}[(1+\ei\cos\nui)\ehatx-\ei\sin\nui\ehaty].
\end{aligned}
\end{eqnarray}
We can then define a new set of unit vectors in terms of the instantaneous velocity of the exploding star. The unit vector pointing in the direction of $\vTwoi$ is
\begin{eqnarray}
    \ehatPar = \frac{1}{\sqrt{1+2\ei\cos\nui + \ei^2}}[(1+\ei\cos\nui)\ehatx-\ei\sin\nui\ehaty].
\end{eqnarray}
The individual velocities can then be expressed as
\begin{eqnarray}
    \vOnei = -\vreli\frac{\mTwoi}{\Mi}\ehatPar,\quad \vTwoi = \vreli\frac{\mOnei}{\Mi}\ehatPar,
\end{eqnarray}
where the relative velocity between the stars is given by
\begin{eqnarray}
    \vreli=\left|\vOnei - \vTwoi\right|=g(\nui, \ei)
    \sqrt{\frac{G\Mi}{\ai}}. \label{appx:vrel}
\end{eqnarray}
The function $g(\nui,\ei)$ represents the deviation from the relative velocity of a circular orbit and is given by
\begin{eqnarray}
    g(\nui,\ei) = \sqrt{\frac{1+2\ei\cos\nui+\ei^2}{1-\ei^2}}.
\end{eqnarray}

It is also useful to define a unit vector perpendicular to $\ehatPar$,
\begin{eqnarray}
    \ehatPrp = \frac{1}{\sqrt{1+2\ei\cos\nui + \ei^2}}[\ei\sin\nui\ehatx+(1+\ei\cos\nui)\ehaty],
\end{eqnarray}    
which satisfies $\ehatPar\times\ehatPrp = \ehatz$. The different unit vectors are then related via
\begin{eqnarray}
\begin{aligned}
    \ehatx = j(\nui, \ei) \ehatPar - h(\nui, \ei) \ehatPrp,\\
    \ehaty = h(\nui, \ei) \ehatPar + j(\nui, \ei) \ehatPrp,
\end{aligned}
\end{eqnarray}
where the functions $h(\nui, \ei)$ and $j(\nui, \ei)$ satisfy that $h(\nui, \ei)^2+j(\nui, \ei)^2=1$ and are defined as
\begin{eqnarray}
    \begin{aligned}
        h(\nui, \ei) = \frac{-\ei\sin\nui}{\sqrt{1+2\ei\cos\nui+\ei^2}}, \\
        j(\nui, \ei) = \frac{1+\ei\cos\nui}{\sqrt{1+2\ei\cos\nui+\ei^2}}.
    \end{aligned}
\end{eqnarray}
In terms of $h(\nui, \ei)$ and $j(\nui, \ei)$ we have that the angle $\tau$, measured counter-clockwise from \ehatPar to \ehatx (see Fig.~\ref{appx:orbitaxes}), satisfies that $\sin\tau=-h(\nui, \ei)$ and $\cos\tau=j(\nui, \ei)$.

For simplicity, we define
\begin{eqnarray}
    \fnui \equiv f(\nui,\ei), \quad \fnuf \equiv f(\nuf,\ef),
\end{eqnarray}
as well as
\begin{eqnarray}
    \gnui \equiv g(\nui,\ei), \quad \hnui \equiv h(\nui,\ei), \quad \jnui \equiv j(\nui,\ei).
\end{eqnarray}

\subsection{Post-explosion semi-major axis}
\begin{figure}
\centering
\includegraphics[width=\columnwidth]{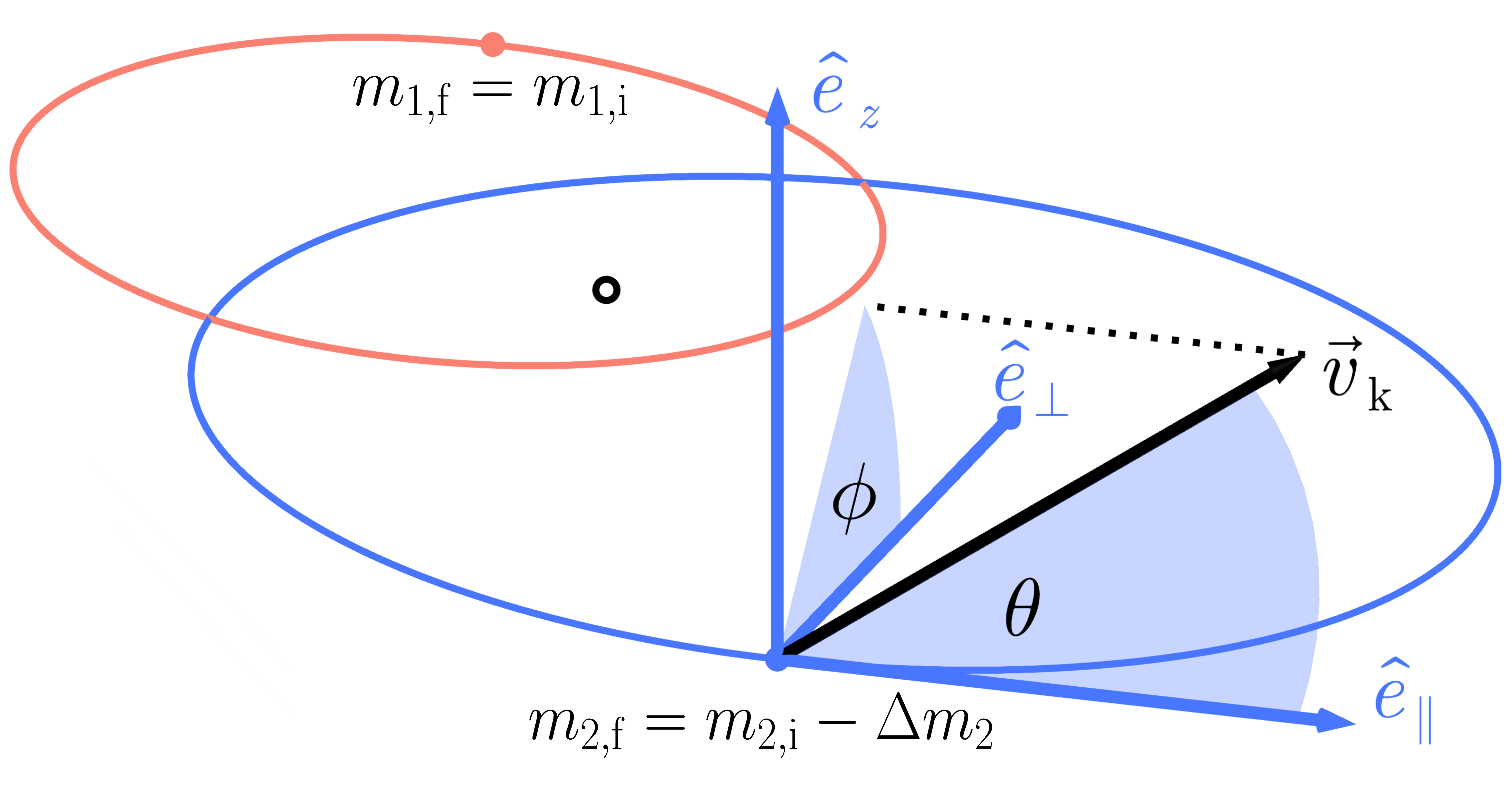}
  \caption{
  \textbf{Illustration of the natal kick}. The angles $\theta$ and $\phi$ define the direction of the kick velocity $\vkickVec$ with respect to the unit vectors $\ehatPar$, $\ehatPrp$ and $\ehatz$.}
     \label{fig:kick_angles}
\end{figure}

We will consider star $2$ undergoes a \ac{SN}, which removes mass leaving a remnant of mass $\mTwof $ while imparting on it a velocity kick $\vkick$. The direction of the kick is defined in terms of a polar angle $\theta$ and an azimuthal angle $\phi$, setting the kick direction with respect to $\ehatPar$, $\ehatPrp$ and $\ehatz$ (see Figure \ref{fig:kick_angles}):
\begin{eqnarray}
\vkickVec = \vkick\vkickHat, \quad \vkickHat=\cos\theta\ehatPar + \sin\theta\cos\phi\ehatPrp + \sin\theta \sin\phi\ehatz.
\end{eqnarray}

For full generality, we will also consider the possibility that the \ac{CC} event removes part of the mass of star $1$, leaving a final mass $\mOnef$ while imparting an impact velocity in the $\ehaty$ direction of magnitude $\vimp$, leaving a system with a final mass $\Mf =\mOnef+\mTwof$. The final velocities of each component are then
\begin{eqnarray}
        \vOnef = \vOnei + \vimp\ehaty,\quad \vTwof = \vTwoi+\vkickVec.
\end{eqnarray}
The final semi-major axis $\af$ can be determined from the total orbital energy in the frame of the new center of mass,
\begin{equation}
    \Ef = -\frac{G\Mf}{2\af},\label{appx:Eorb}
\end{equation}
which can be computed from the separation and velocities post-explosion (with respect to the new center of mass). The center of mass velocity post-explosion is given by
\begin{eqnarray}
\vCMf = \frac{\mOnef \vOnef+\mTwof \vTwof}{\Mf},\label{appx:vcm}
\end{eqnarray}
from which the individual velocities in the reference frame of the new center of mass are
\begin{eqnarray}
\begin{aligned}
\vCMOnef = \vOnef - \vCMf = \frac{\mTwof }{\Mf}(\vOnef-\vTwof),\\
\vCMTwof = \vTwof - \vCMf = \frac{\mOnef }{\Mf}(\vTwof-\vOnef).
\end{aligned}\label{appx:vcms}
\end{eqnarray}
The final orbital energy can then be computed by considering the instantaneous orbital separation remains fixed,
\begin{equation}
\Ef = -\frac{G\mOnef \mTwof }{\ai\fnui} + \frac{\mOnef \mTwof }{2\Mf }\left(\vTwof-\vOnef\right)^2,\label{appx:Ef}
\end{equation}
where to compute the kinetic energy term it is convenient to write
\begin{eqnarray}
    \vTwof-\vOnef = \vreli\left(\ehatPar-\beta \ehaty+\alpha\vkickHat\right),\quad\beta \equiv \frac{\vimp}{\vreli},\quad\alpha\equiv\frac{\vkick}{\vreli}.\label{appx:diffvf}
\end{eqnarray}
Expanding $\ehaty$ and $\vkickHat$ in terms of $\ehatPar$, $\ehatPrp$ and $\ehatz$, it is straightforward to show that
\begin{eqnarray}
\begin{aligned}
    \frac{\left(\vTwof-\vOnef\right)^2}{\vreli^2}=1+\alpha^2+\beta^2+2[\alpha\cos\theta -\hnui\beta(1+\alpha\cos\theta) \\ -\jnui\beta\alpha\sin\theta\cos\phi].
\end{aligned}
\end{eqnarray}
Combining Equations (\ref{appx:vrel}) and (\ref{appx:Ef}) we can express the total energy as
\begin{eqnarray}
    \Ef = -\frac{G\mOnef \mTwof }{\fnui \ai}\left(1-\frac{\xi}{2}\right),
\end{eqnarray}
where
\begin{eqnarray}
    \xi \equiv \fnui \gnui^2\frac{\Mi}{\Mf }\frac{\left(\vTwof-\vOnef\right)^2}{\vreli^2}.
\end{eqnarray}
This also implies that the orbit becomes unbound if $\xi>2$. Equating the last expression to the final orbital energy given in Equation (\ref{appx:Eorb}) we obtain the post-explosion semi-major axis:
\begin{eqnarray}
    \af=\frac{\fnui\ai}{2-\xi}.\label{appx:af}
\end{eqnarray}
\subsection{Post-explosion eccentricity}
Similar to the derivation of the post-explosion semi-major axis, the post-explosion eccentricity can be computed in terms of the final orbital angular momentum,
\begin{equation}
    \Lf = \mOnef \mTwof \sqrt{\frac{G\af(1-\ef ^2)}{\Mf}}.\label{appx:Lf}
\end{equation}
The orbital angular momentum in the new center of mass frame is also given by
\begin{equation}
    \LfVec = \mOnef \rCMOnef\times\vCMOnef +  \mTwof \rCMTwof\times\vCMTwof,
\end{equation}
where $\rCMOnef$ and $\rCMTwof$ are vectors pointing from the new center of mass to each component,
\begin{equation}
    \rCMOnef = \fnui \ai \frac{\mTwof }{\Mf }\ehaty,\quad \rCMTwof = -\fnui \ai \frac{\mOnef }{\Mf }\ehaty.\label{appx:rcm}
\end{equation}

Combining the previous two Equations together with Equations (\ref{appx:vcms}) and (\ref{appx:diffvf}) we find that
\begin{equation}
    \LfVec = -\fnui \ai\frac{\mOnef \mTwof }{\Mf}\vreli\left[\ehaty\times (\ehatPar +\alpha \vkickHat)\right]. \label{appx:Lder}
\end{equation}
The magnitude of the vector in brackets is straightforward to determine by expressing it in terms of $\ehatPar$, $\ehatPrp$ and $\ehatz$, resulting in
\begin{eqnarray}
\begin{aligned}
    \left[\ehaty\times (\ehatPar +\alpha \vkickHat)\right]^2 = \alpha^2\sin^2\theta\sin^2\phi + [\hnui \alpha \sin\theta\cos\phi\qquad \\- \jnui (1+\alpha\cos\theta)]^2.
\label{appx:Lnorm}
\end{aligned}
\end{eqnarray}
Combing Equations (\ref{appx:Lder}) and (\ref{appx:vrel}) we find that
\begin{eqnarray}
    \Lf^2 = \fnui \ai G
    \frac{(\mOnef \mTwof )^2}{\Mf }\eta.\label{appx:Lf2}
\end{eqnarray}
where
\begin{eqnarray}
    \eta\equiv \fnui \gnui^2\frac{\Mi}{\Mf }\left[\ehaty\times (\ehatPar +\alpha \vkickHat)\right]^2.
\end{eqnarray}
    
Combining Equations (\ref{appx:af}), (\ref{appx:Lf}) and (\ref{appx:Lf2}) we obtain the post-explosion orbital eccentricity:
\begin{eqnarray}
    \ef  = \sqrt{1+(\xi-2)\eta}.
\end{eqnarray}
\subsection{Center of mass velocity}

The velocity of the center of mass can be determined from Equation (\ref{appx:vcm}), but it can also be determined in a straightforward way by accounting for the change in momentum of the binary components. If we write the center of mass velocity as
\begin{eqnarray}
    \vCMf = \vCMPar\ehatPar + \vCMPrp\ehatPrp + \vCMz\ehatz,
\end{eqnarray}
then the individual components of the velocity are given by
\begin{eqnarray}
    \begin{aligned}
    \vCMPar =& \frac{1}{\Mf }\left[
    \frac{\vreli}{\Mi}
    (\mOnei\mTwof  - \mOnef \mTwoi)
    \right.\\
    &\left. \mTwof \vkick\cos\theta + \hnui \mOnef \vimp \vphantom{\frac{m_1}{M}}\right],\\
    \vCMPrp =& \frac{1}{\Mf }\left(\mTwof \vkick\sin\theta\cos\phi + \jnui \mOnef \vimp\right),\qquad\qquad\quad\;\\
    \vCMz =& \frac{\mTwof \vkick\sin\theta\sin\phi}{\Mf }.\qquad\qquad\qquad\qquad\qquad\;\;
    \end{aligned}
\label{appx:vcmcomp}
\end{eqnarray}
We note that this is the final velocity of the center of mass of the binary in a reference frame where the binary was static prior to the explosion, while for comparison to an observed system we need to account for any pre-collapse velocity.

\subsection{Observer frame of reference}
\label{appx:observer_frame_of_ref}

\begin{figure*}[hbt!]
\centering
\includegraphics[width=2\columnwidth]{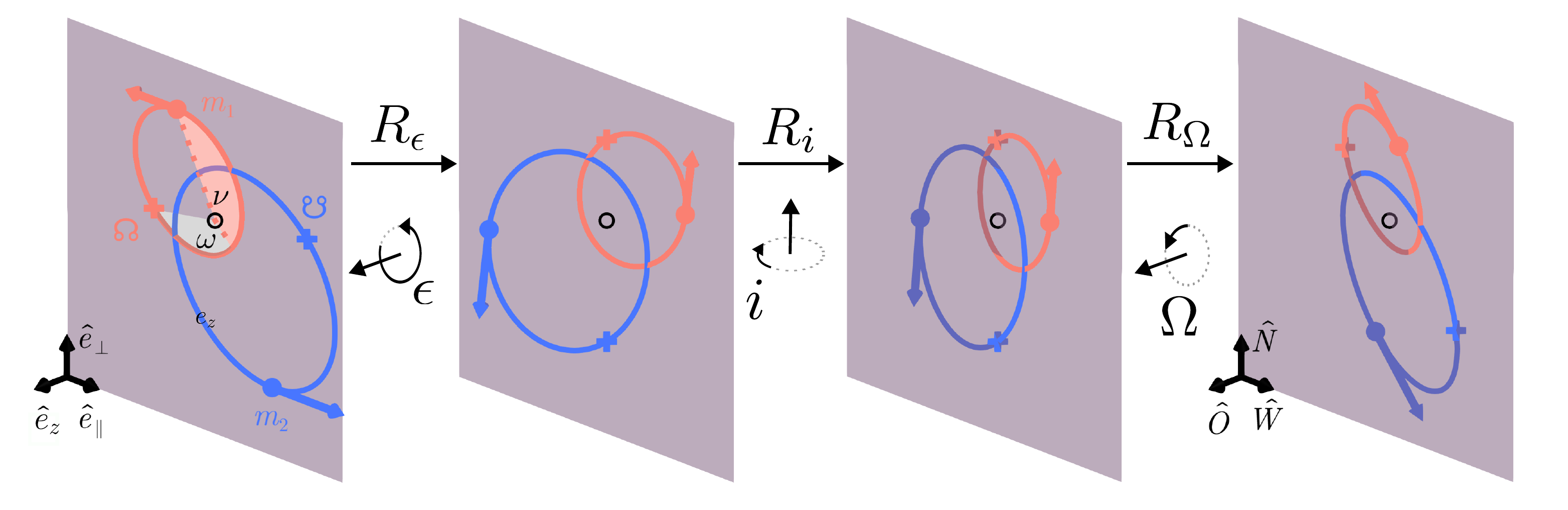}
  \caption{
  \textbf{The series of rotations to convert from the frame of reference of the binary to that of the observer.} 
  The binary frame of reference is spanned by the unit vectors $\ehatPar$, $\ehatPrp$ and $\ehatz$, while the observer frame of reference is defined by $\hat{N}$, $\hat{W}$, and $\hat{O}$.
  The left-most figure shows the pre-explosion system including the position of the ascending node for $m_1$ and the descending node for $m_2$, indicated with pluses. 
  The following panels apply the rotation matrices $R_\epsilon$, $R_i$ and $R_\Omega$. 
  For reference, a transparent plane is shown that remains fixed during rotations, with its edges aligned with $\ehatPar$ and $\ehatPrp$ at the leftmost panel (darker shading indicates the orbit is behind the plane).
  }
     \label{appx:rotations}
\end{figure*}

We also need to account for the relationship between the direction of motion of the system and the Keplerian elements, including the inclination $\inc$, the longitude of the ascending node $\Omega$, and the argument of periastron $\omega$ of the companion. 
We take the convention here that the ascending node corresponds to the intersection point of the orbit with the celestial plane at which the star is moving away from the observer. 
We derive the expected properties that would be visible from a reference frame centered on the observer, with unit vectors pointing in the directions of celestial north $\hat{N}$, celestial west $\hat{W}$, and towards the observer $\hat{O}$ (these then satisfy $\hat{W}\times\hat{N}=\hat{O}$). 
This basis is identical (up to reflection) of the $\delta-\alpha-r$ basis used above for the velocity components, but is used here to make it easier to visualize the rotations.

If we consider a vector
\begin{eqnarray}
    \vec{b}=\bPar\ehatPar + \bPrp \ehatPrp + b_z \ehatz = \bW\hat{W} + \bN \hat{N} + \bO \hat{O},
\end{eqnarray}
then the components of $\vec{b}$ in the observer frame can be determined by a series of rotations,
\begin{eqnarray}
    \left(\begin{matrix}\bW \\ \bN \\ \bO\end{matrix}\right) = R_\Omega R_i R_\epsilon\left(\begin{matrix}\bPar \\ \bPrp \\ b_z\end{matrix}\right).
\label{appx:rots}
\end{eqnarray}
The rotation matrix $R_\epsilon$ is meant to put the ascending node of star 1 in the direction $(0,1,0)$, and is given by
\begin{eqnarray}
    R_\epsilon=\left(
    \begin{matrix}
        \cos\epsilon & -\sin\epsilon & 0 \\
        \sin\epsilon &  \cos\epsilon & 0 \\
        0 & 0 & 1
    \end{matrix}
    \right).
\end{eqnarray}
where $\epsilon=\omega + \nu - \tau$, 

\begin{eqnarray}
    \tau = \begin{cases}
    \;\;\,\arccos \jnu & 0 \le \nu < \pi \\
        -\arccos \jnu & \pi \le \nu < 2\pi
    \end{cases}.
\end{eqnarray}

The following rotation matrix, $R_i$, corrects for the inclination and sets the direction $(0,0,1)$ to point toward the observer:
\begin{eqnarray}
    R_i = \left(
    \begin{matrix}
        \cos \inc & 0 & -\sin \inc \\
        0 & 1 & 0 \\
        \sin \inc & 0 & \cos \inc
    \end{matrix}
    \right).
\end{eqnarray}
We emphasize that the inclination here is defined on the range $[0,\pi]$, to account for all inclinations in which the star is moving away from the observer at the ascending node.

Finally, the rotation $R_\Omega$ rotates the projected orbit in the plane of the sky by $\Omega$ so that the direction $(0,1,0)$ points toward the ascending node,
\begin{eqnarray}
    R_\Omega =\left(
    \begin{matrix}
        \cos\Omega & -\sin\Omega & 0 \\
        \sin\Omega & \cos\Omega & 0 \\
        0 & 0 & 1
    \end{matrix}\right).        
\end{eqnarray}
Note that we have omitted time-dependent subscripts here as the rotation can be applied to both the pre- and post-collapse binary. The center of mass velocity of the binary in the observer frame can then be obtained by applying Equation (\ref{appx:rots}) to the components of $\vCMf$.

\subsection{Post-explosion inclination}
To determine the inclination of the resulting binary we need to compute the angle between the orbital angular momentum post-explosion and the unit vector $\hat{O}$. Starting from Equation \ref{appx:Lder} we have that the unit vector pointing in the direction of the final orbital angular momentum is
\begin{eqnarray}
    \LfHat = \frac{-\ehaty\times(\ehatPar + \alpha \vkickHat)}{|\ehaty\times(\ehatPar + \alpha \vkickHat)|},
\end{eqnarray}
which has components
\begin{eqnarray}
    \begin{aligned}
        \LfHat\cdot\ehatPar &=& -\frac{\jnui \alpha \sin\theta\sin\phi}{|\ehaty\times(\ehatPar + \alpha\vkickHat)|} \qquad\qquad\qquad\;\;\\
        \LfHat\cdot\ehatPrp &=& \frac{\hnui \alpha \sin\theta\sin\phi}{|\ehaty\times(\ehatPar + \alpha\vkickHat)|}
        \qquad\qquad\qquad\;\\
        \LfHat\cdot\ehatz &=& -\frac{\hnui \alpha \sin\theta\cos\phi - \jnui (1+\alpha\cos\theta)}{|\ehaty\times(\ehatPar + \alpha\vkickHat)|},
    \end{aligned}
\end{eqnarray}
with the denominator given by Equation (\ref{appx:Lnorm}). The full components of the angular momentum in the base given by $\hat{W}$, $\hat{N}$ and $\hat{O}$ can then be derived using Equation (\ref{appx:rots}). With the component of $\LfHat$ that points in the $\hat{O}$ direction we can directly extract the post-explosion inclination:
\begin{eqnarray}
    \incf = \arccos(\LfHat\cdot \hat{O})
\end{eqnarray}

\subsection{Post-explosion longitude of the ascending node}
\begin{figure}
\centering
\includegraphics[width=\columnwidth]{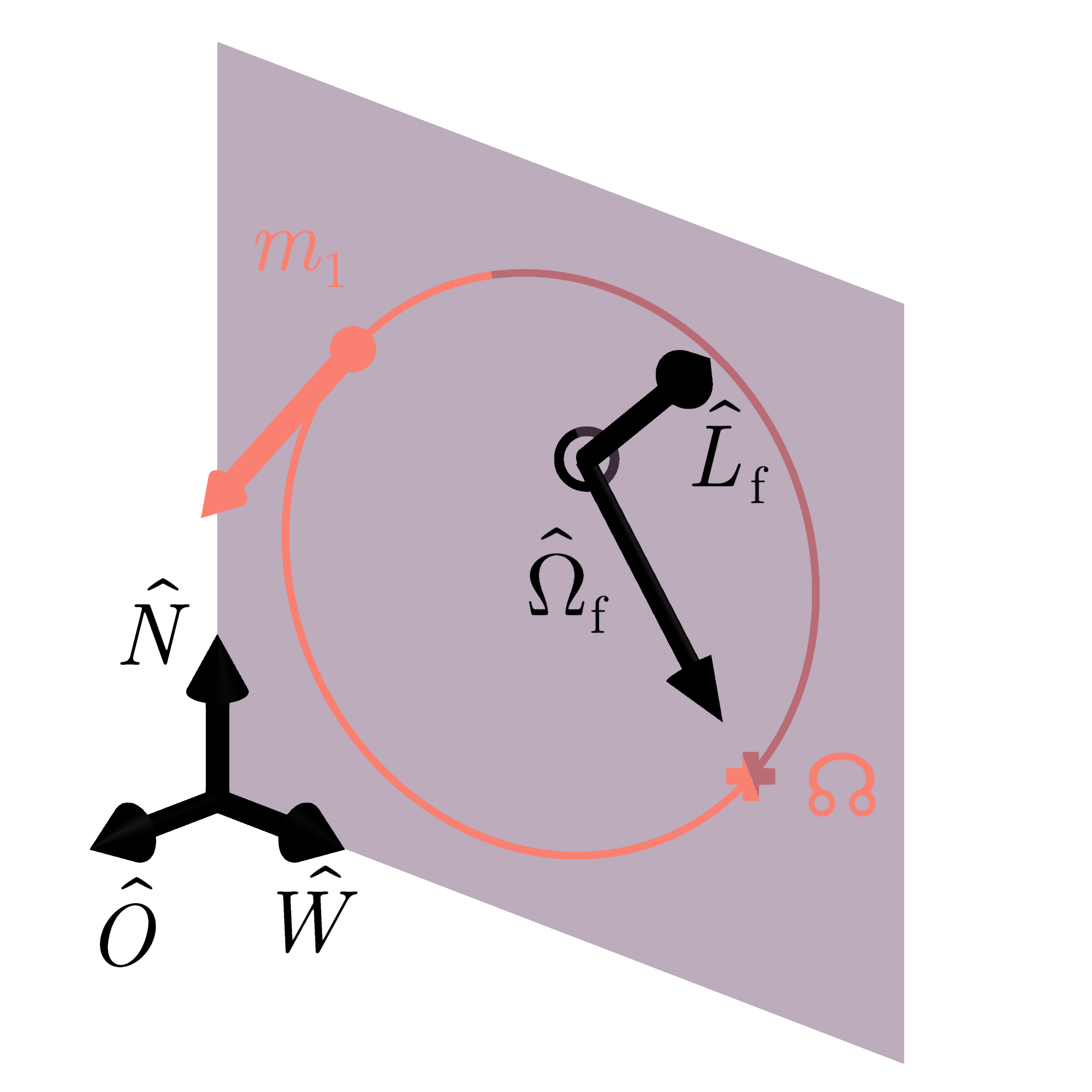}
  \caption{
  \textbf{Orientation of the orbital angular momentum in the observer reference frame.}
  The orbital angular momentum vector, $\LfHat$, and a vector pointing from the center of mass of the binary to the ascending node, $\OmegafHat$, orient the post-explosion orbit of the visible star. 
  Both $\LfHat$ and $\hat{O}$ are perpendicular to $\OmegafHat$, but are not perpendicular to each other. 
  In the particular case shown here $\OmegafHat\cdot \hat{W}>0$, such that $\Omegaf=2\pi-\arccos(\OmegafHat\cdot \hat{N})$.
  }
     \label{fig:asc_node}
\end{figure}

The longitude of the ascending node is the angle, measured counterclockwise on the plane of the sky, between the direction $\hat{N}$ and the vector pointing from the center of mass towards the ascending node. Such a unit vector can be taken by the cross product between $\LfHat$ and $\hat{O}$, both of which are perpendicular to the vector pointing towards the ascending node. Calling this unit vector $\OmegafHat$ and noting that the ascending node is the one where the object is moving away from the observer, we have that (see Figure \ref{fig:asc_node})
\begin{eqnarray}
    \OmegafHat = \frac{\LfHat\times \hat{O}}{|\LfHat\times \hat{O}|}.
\end{eqnarray}
The longitude of the ascending node can then be computed as
\begin{eqnarray}
    \Omegaf = \begin{cases}
        \arccos\left(\OmegafHat\cdot \hat{N}\right) & \OmegafHat\cdot \hat{W} \le 0 \\
        2\pi - \arccos\left(\OmegafHat\cdot \hat{N}\right) & \OmegafHat\cdot \hat{W} > 0
    \end{cases}.
\end{eqnarray}

\subsection{Post-explosion true anomaly}

We next calculate the true anomaly $\nuf$ immediately after the explosion, which we will use to calculate the post-explosion argument of periastron $\omegaf$. The distance between the binary components is the same right before and right after the explosion, such that (see Equation~\ref{eq:fnu}):
\begin{equation}
    \fnui\ai = \fnuf\af.
\end{equation}
The cosine of the final true anomaly is then
\begin{equation}
    \cos\nuf = \frac{1}{\ef}\left[\frac{\af(1 - \ef ^2)}{\fnui\ai} - 1\right].
\end{equation}
The actual value then depends on whether we are moving away ($0\le \nuf<\pi$) or towards ($\pi \le \nuf < 2\pi$) periastron. This can be determined with the $\ehaty$ component of $\vCMOnef$, as this is still the axis connecting the two components.
From Equation (\ref{appx:vcmcomp}) we have that

\begin{figure*}
\centering
\includegraphics[width=2\columnwidth]{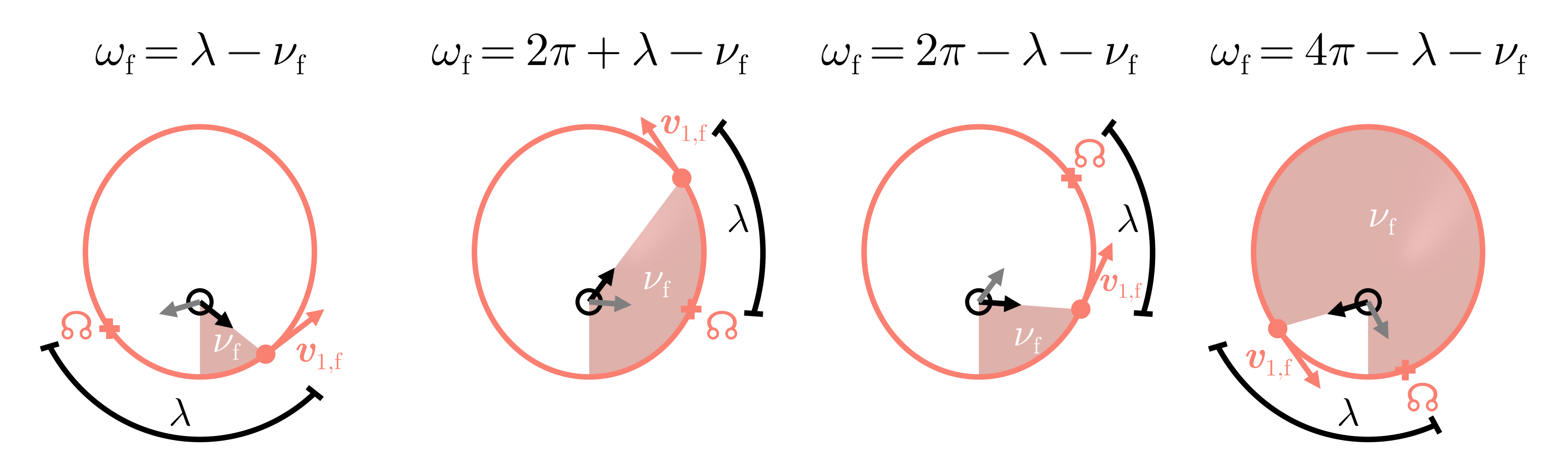}
\caption{
\textbf{Illustration of the different special cases to compute the argument of periastron for the post-explosion orbit of the visible star.} 
See Equation \ref{equ:arg_periastron} for reference.
The black arrow shows the vector $\hat{y}$ while the gray arrow shows $\OmegafHat$, which point from the center of mass to the visible star and the ascending node respectively.
}
\label{fig:arg_periastron}
\end{figure*}    

\begin{equation}
\vCMOnef\cdot\ehaty = \vimp - \hnui\left(\frac{\mTwoi}{\Mi}\vreli + \vCMPar\right) - \jnui \vCMPrp.
\end{equation}
Using this, the true anomaly post-explosion is given by:
\begin{eqnarray}
\nuf = \begin{cases}
    \displaystyle\arccos\left\{\frac{1}{\ef }\left[\frac{\af(1-\ef ^2)}{\fnui \ai}-1\right]\right\} & \vCMOnef \cdot \ehaty > 0 \\
    \displaystyle 2\pi - \arccos\left\{\frac{1}{\ef }\left[\frac{\af(1-\ef ^2)}{\fnui \ai}-1\right]\right\} &  \vCMOnef \cdot \ehaty < 0
    \end{cases}.
\end{eqnarray}
\subsection{Post-explosion argument of periastron}

Given that we know a unit vector pointing from the center of mass to star 1, and a unit vector pointing towards the ascending node, we can determine the angle between these two vectors,
\begin{eqnarray}
    \lambda = \arccos\left(\hat{y}\cdot\OmegafHat\right).
\end{eqnarray}
In order to determine the argument of periastron after explosion $\omegaf$, we need to combine $\nuf$ and $\lambda$ in a smart way, depending on the relative location between star 1 and the ascending node in the post-explosion orbit. The full set of conditions is as follows
\begin{eqnarray}
\omegaf = 
\begin{cases}
    \lambda-\nuf & (\ehaty\times\OmegafHat)\cdot\LfHat\le0 \;\wedge\; \lambda > \nuf \\
    2\pi+\lambda-\nuf & (\ehaty\times\OmegafHat)\cdot\LfHat\le0 \;\wedge\; \lambda \le \nuf \\
    2\pi-\lambda-\nuf & (\ehaty\times\OmegafHat)\cdot\LfHat>0 \;\wedge\; \lambda+\nuf \le 2\pi \\
    4\pi-\lambda-\nuf & (\ehaty\times\OmegafHat)\cdot\LfHat>0 \;\wedge\; \lambda+\nuf > 2\pi 
\end{cases}.
\label{equ:arg_periastron}
\end{eqnarray}
Each of these cases is illustrated in Figure \ref{fig:arg_periastron}.

\section{Injection analysis for the eccentric system}\label{appx:injection_analysis_extras}

\begin{figure*}[htp]
  \centering
  \subfigure[
  Initially eccentric binary, without a resolved orbit.
  ]{\includegraphics[scale=.85]{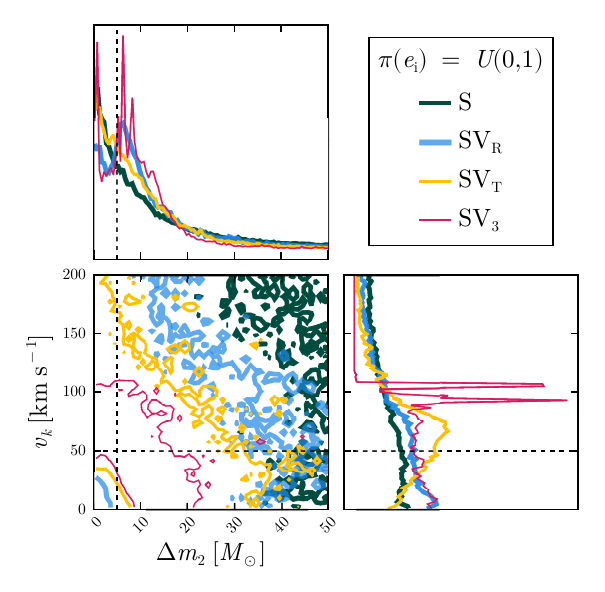}}
  \hfill
  \subfigure[
  Initially eccentric binary, with a resolved orbit.
  ]{\includegraphics[scale=.85]{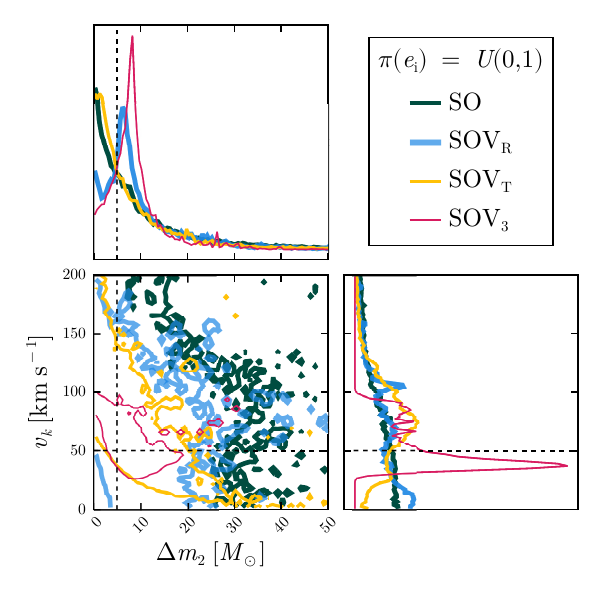}}
  \caption{
  \textbf{The impact of including different observations in a mock eccentric binary.}
  Posteriors on the natal kick (\vkick) and mass loss (\dMTwo), when the various observational categories are included. 
  Observational categories include spectroscopy (S), resolved astrometric orbit (O), and constraints on any of radial velocity (V$_\mathrm{R}$), transverse velocities (V$_\mathrm{T}$), or all three velocity components (V$_\mathrm{3}$).
  All categories include spectroscopy (see Sec.~\ref{sec:observational_categories} for justification).
  Categories without (with) the astrometric orbit are shown on the left (right).
  Colors correspond to differences in the available velocity constraints: 
  when there are no velocity constraints (e.g., due to no host association; \modelS, \modelSO: dark green), 
  when there are measured \acp{RV} for the source and host (\modelSVr, \modelSOVr: blue),
  when there are measured transverse velocities for the source and host (\modelSVt, \modelSOVt: yellow),
  and when there are measured velocities in all three directions for the source and host (\modelSVthree, \modelSOVthree: red).
  All results here apply for the eccentric mock system; the results for the circular mock system are shown in Fig.~\ref{fig:injection_corner_plots_obs_categories_circ}.
  Black dashed lines show the true values.
  Contours capture the 90\% confidence interval of the 2D distributions.
  }
\label{fig:injection_corner_plots_obs_categories_ecc}
\end{figure*}

In Fig.~\ref{fig:injection_corner_plots_obs_categories_ecc}, we provide the full corner plot that results from the injection analyses binary, while varying which observations are included.
The plot is the same as Fig.~\ref{fig:injection_corner_plots_obs_categories_circ} except this one is for the initially eccentric mock system.
All posteriors are substantially broader than in the case of the circular mock binary.
However, for all the models with any velocity constraints (namely, all except \modelS and \modelSO), we can still rule out the 0~\kms natal kick - 0~\Msun corner associated with direct collapse \acp{SN} at the 90\% confidence level.

%
%
%
%


\section{Full corner plots for VFTS 243}\label{appx:vfts243_cornerplots}
    
\begin{figure*}
\centering
\includegraphics[width=\textwidth]{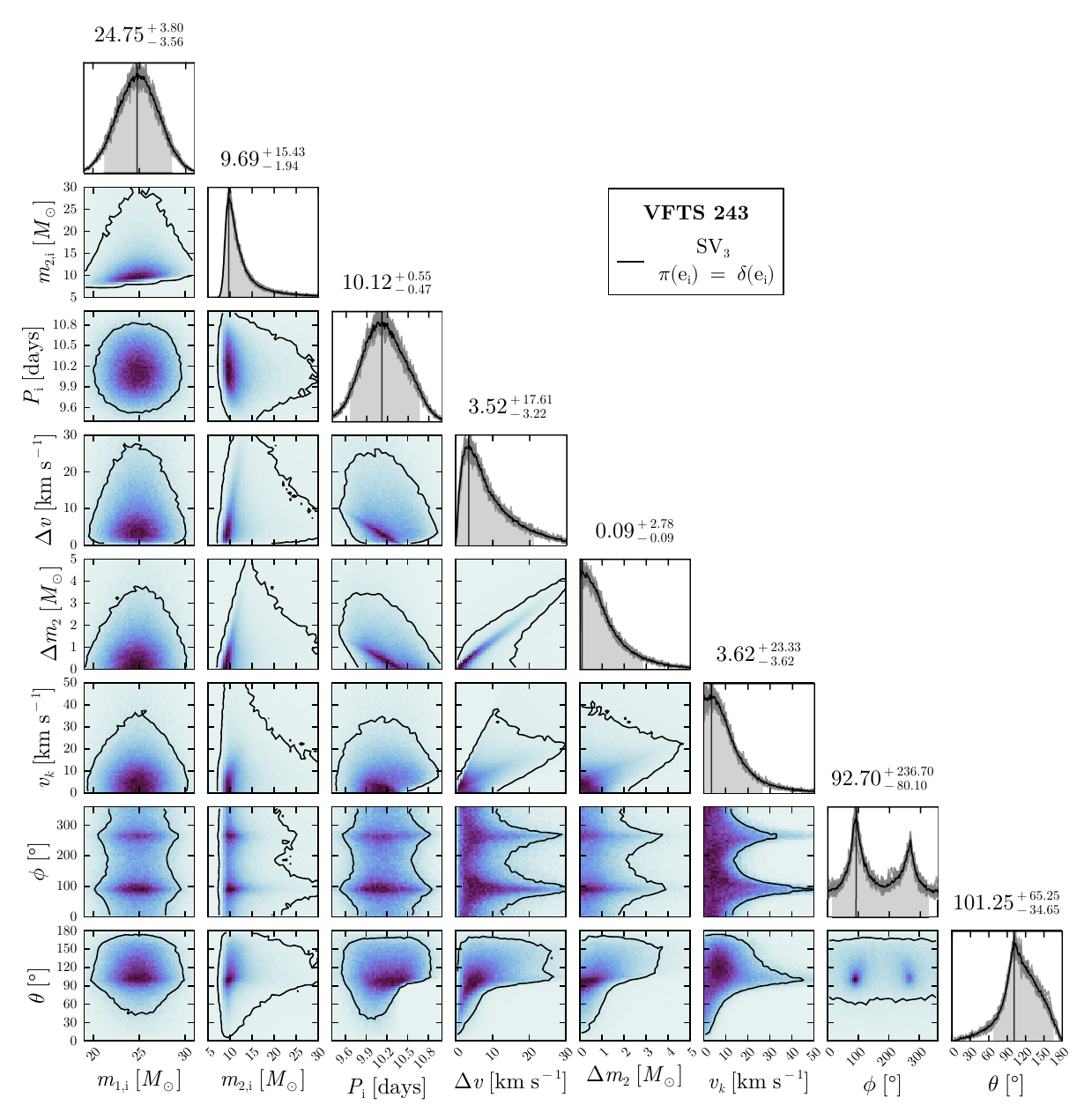}
\caption{
\textbf{Corner plot for VFTS 243, using the circular prior.} 
Posteriors shown here include the initial masses and orbital period, the post-\ac{SN} system velocity relative to the birth velocity, and the parameters that describe the \ac{SN} mass loss and natal kick.
The observing model is \modelSVthree, which is the best case available for this system given that it is too far to resolve the orbit. 
}
\label{fig:vfts243_derivables_circ}
\end{figure*}

\begin{figure*}
\centering
\includegraphics[width=\textwidth]{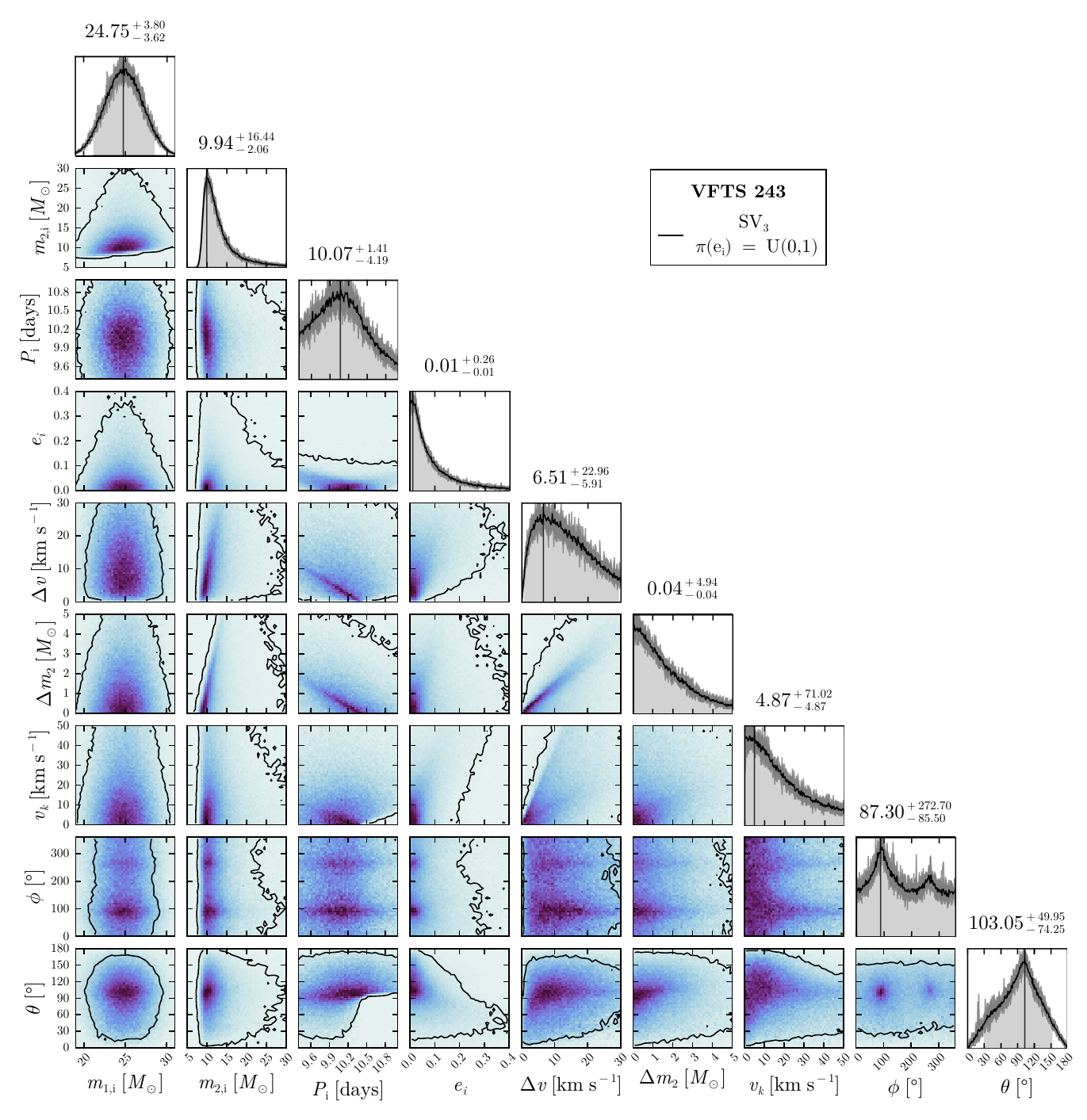}
\caption{
\textbf{Corner plot for VFTS 243, using the eccentric prior.} Posteriors shown here include the initial masses, orbital period, and eccentricity, the post-\ac{SN} system velocity relative to the birth velocity, and the parameters that describe the \ac{SN} mass loss and natal kick.
The observing model is \modelSVthree, which is the best case available for this system given that it is too far to resolve the orbit. 
}
\label{fig:vfts243_derivables_ecc}
\end{figure*}

In Figs.~\ref{fig:vfts243_derivables_circ} and \ref{fig:vfts243_derivables_ecc}, we show the corner plots for the VFTS 243 analysis, for a broader set of the parameters of interest. 
Both plots use the best case observing scenario \modelSVthree; In Fig.~\ref{fig:vfts243_derivables_circ}, we used the circular prior $\pi(\ei) = \delta(\ei)$, and in  Fig.~\ref{fig:vfts243_derivables_ecc}, we used the eccentric prior $\pi(\ei) = U(0,1)$.

\end{appendix}

\label{LastPage}
\end{document}